\documentclass[sigconf,10pt,nonacm]{acmart}

\setcopyright{none}

\usepackage{balance}

\def\BibTeX{{\rm B\kern-.05em{\sc i\kern-.025em b}\kern-.08emT\kern-.1667em\lower.7ex\hbox{E}\kern-.125emX}}

\copyrightyear{2019}
\acmYear{2019}
\setcopyright{acmcopyright}
\acmConference[PODS'19]{38th ACM SIGMOD-SIGACT-SIGAI Symposium on Principles of Database Systems}{June 30-July 5, 2019}{Amsterdam, Netherlands} \acmBooktitle{38th ACM SIGMOD-SIGACT-SIGAI Symposium on Principles of Database Systems (PODS'19), June 30-July 5, 2019, Amsterdam, Netherlands} \acmPrice{15.00}
\acmDOI{10.1145/3294052.3319679} \acmISBN{978-1-4503-6227-6/19/06}

\usepackage{amsmath,amsthm,bm,multicol,graphicx}
\usepackage[noend]{algpseudocode}
\usepackage{algorithmicx}
\usepackage{algorithm}
\usepackage{comment,url}
\usepackage{ifthen}

\newboolean{fullversion}
\setboolean{fullversion}{true} %

\newtheorem{theorem}{Theorem}[section]

\newtheorem{proposition}[theorem]{Proposition}

\newcommand{\bfA}{\mathbf{A}}
\newcommand{\bfB}{\mathbf{B}}
\newcommand{\bfC}{\mathbf{C}}
\newcommand{\bfG}{\mathbf{G}}
\newcommand{\bfH}{\mathbf{H}}
\newcommand{\bfI}{\mathbf{I}}
\newcommand{\bfa}{\mathbf{a}}
\newcommand{\bfb}{\mathbf{b}}
\newcommand{\bfi}{\mathbf{i}}

\newcommand{\bfq}{\mathbf{q}}

\newcommand{\bft}{\mathbf{t}}
\newcommand{\caA}{\mathcal{A}}
\newcommand{\caB}{\mathcal{B}}
\newcommand{\caD}{\mathcal{D}}
\newcommand{\caE}{\mathcal{E}}

\newcommand{\caI}{\mathcal{I}}

\newcommand{\caM}{\mathcal{M}}

\newcommand{\bme}{{\bm e}}
\newcommand{\bmm}{{\bm m}}

\newcommand{\ar}{\mathrm{ar}}
\newcommand{\bbN}{\mathbb{N}}
\newcommand{\bbR}{\mathbb{R}}

\newcommand{\E}{\mathop{\mathbf{E}}}

\newcommand{\set}[1]{\{#1\}}
\newcommand{\multiset}[1]{\{\{#1\}\}}
\allowdisplaybreaks

\title{Testability of Homomorphism Inadmissibility:\\Property Testing Meets Database Theory}
\author{Hubie Chen}
\authornote{Acknowledges the support of Spanish Project TIN2017-86727-C2-2-R.}
\affiliation{Dept.\ of Informatics\\
King's College London\\
London, United Kingdom\\
}
\email{hubie@kcl.ac.uk}

\author{Yuichi Yoshida}
\authornote{Supported by JSPS KAKENHI Grant Number JP17H04676.}
\orcid{0000-0001-8919-8479}
\affiliation{National Institute of Informatics \\
  Tokyo, Japan\\
}
\email{yyoshida@nii.ac.jp}

\renewcommand{\shortauthors}{Chen and Yoshida}

\begin{document}

\begin{abstract}
In this paper, we utilize the perspective of property testing
to consider the testability of relational database queries.
A primary motivation is the desire to avoid reading
an entire database to decide a property thereof.
We focus on conjunctive queries,
which are the most basic and heavily studied database queries.
Each conjunctive query can be represented as a
relational structure $\mathbf{A}$ such that
deciding if the conjunctive query is satisfied by
a relational structure $\mathbf{B}$ is equivalent
to deciding if there exists a homomorphism from $\mathbf{A}$ to $\mathbf{B}$.
We phrase our results in terms of homomorphisms.

Precisely, we study,
for each relational structure $\mathbf{A}$,
the testability of homomorphism inadmissibility from $\mathbf{A}$.
We consider algorithms that have
oracle access to an input relational structure $\mathbf{B}$
and that distinguish, with high probability,
the case where there is no homomorphism from $\mathbf{A}$ to $\mathbf{B}$,
from the case where one needs to remove
a constant fraction of tuples from $\mathbf{B}$ in order to suppress all such homomorphisms.
We provide a complete characterization of the structures
$\mathbf{A}$ from which one can test homomorphism inadmissibility with one-sided error by making a constant number of queries to $\mathbf{B}$.
Our characterization shows that
homomorphism inadmissibility from $\mathbf{A}$ is constant-query testable with one-sided error if and only if the core of $\mathbf{A}$ is $\alpha$-acyclic.
We also show that the injective version of the problem is constant-query testable with one-sided error if $\mathbf{A}$ is $\alpha$-acyclic;
this result generalizes existing results for testing subgraph-freeness in the general graph model.
\end{abstract}


\maketitle

\section{Introduction}

\subsection{Background}

{\bf Property testing.}
Traditional notions of algorithmic efficiency,
such as that of polynomial time,
permit each input to be read in its entirety.
However, the modern necessity of performing
computations on large bodies of data
poses situations where inputs are so large
that it is prohibitively expensive to fully read them.
These situations motivate the study of algorithms
that only read a part of each input.
\emph{Property testing} provides a framework wherein
it is possible to present such algorithms.

In property testing,
the input is typically read via random access queries,
and an algorithm's desideratum is to distinguish between
inputs that satisfy a property and
inputs that are \emph{$\epsilon$-far} from satisfying the property.
Intuitively speaking, an object is called \emph{$\epsilon$-far} from satisfying a property if
one needs to modify more than an $\epsilon$-fraction of the object
in order for it satisfy the property;
the exact definition depends on the context.
The restriction on the allowed inputs
leads to the possibility of algorithms that do not read the entire input.

A diversity of combinatorial objects have been studied
in property testing,
including functions~\cite{Blum:1993cn,Rubinfeld:1996um}, graphs~\cite{Goldreich:1998wa,Goldreich:2002bn}, and constraint satisfaction problems (CSPs)~\cite{Alon:2003jp,Yoshida:2014zn};
many properties have been shown to be testable with a constant number of queries, a most positive type of result.
Moreover, general characterizations of constant-query testable properties have been achieved in several settings, including affine-invariant properties of functions over finite fields~\cite{Yoshida:2014tq,Yoshida:2016zz}, properties on dense graphs~\cite{Alon:2009gn}, and assignment testing of CSPs~\cite{Bhattacharyya:2013fa,Chen:2016wf}.
See the surveys~\cite{ron2010algorithmic,rubinfeld2011survey} and the book~\cite{goldreich2010book}.

{\bf Database theory, conjunctive queries,
and relational homomorphisms.}
A central, fundamental computational task
in the field of databases is the evaluation of
a \emph{query} on a \emph{relational database}.
In this article, we focus on \emph{conjunctive queries},
the most basic and heavily studied class of database queries~\cite{SchweikardtSchwentickSegoufin09-querylanguages}.
While there are numerous equivalent definitions,
a \emph{conjunctive query} may be defined
as a first-order logical sentence
built from predicate applications, conjunction, and existential quantification.
(We focus here on \emph{Boolean} conjunctive queries,
which are conjunctive queries without so-called free variables
and which evaluate to either true or false on a given database.)
As is usual in database theory,
we formalize a relational database
as a \emph{relational structure}, defined as an indexed
set of relations, all over the same domain.
As an example, consider the evaluation
of the conjunctive query
$\psi = \exists v_1 \exists v_2 \exists v_3
(E(v_1,v_2) \wedge E(v_2, v_3) \wedge E(v_3, v_1))$
on a simple undirected graph, whose edges
are represented by a symmetric binary relation;
this conjunctive query evaluates to true if and only if
the graph contains a triangle.

A classical result in database theory~\cite{Chandra:1977ez}
is that
each conjunctive query $\phi$ can be naturally passed
to a relational structure $\bfA[\phi]$,
in a way such that, for any relational structure $\bfB$,
the query $\phi$ evaluates to true on $\bfB$
if and only if there exists a
\emph{homomorphism} from $\bfA[\phi]$ to $\bfB$.
Moreover, this passage can be naturally inverted.
Hence, the problem of conjunctive query evaluation
is equivalent to the \emph{relational homomorphism problem}
where, given
a pair of relational structures, the goal is to decide if there exists
a homomorphism from the first structure to the second.

Let us briefly present the relevant definitions and notions.
Define a \emph{relational structure}
$\bfA = (A; R_1^\bfA,\ldots,R_k^\bfA)$
to consist of a finite set $A$,
called the \emph{domain},
and relations $R_i^{\bfA}~(i \in \set{1,\ldots,k})$.
Each relation symbol $R_i$ has an associated arity $\ar(R_i)$,
and each $R_i^{\bfA}$ is a relation of arity $\ar(R_i)$ over $A$,
that is, a collection of tuples
where each has length $\ar(R_i)$ and has entries drawn from $A$.
When
$\bfA = (A; R_1^\bfA,\ldots,R_k^\bfA)$
and $\bfB = (B; R_1^\bfB,\ldots,R_k^\bfB)$
are structures defined on the same symbols
(that is, having the same \emph{vocabulary}),
a \emph{homomorphism} from $\bfA$ to $\bfB$
is a mapping $h: A \to B$ such that
for every $i \in \set{1,\ldots,k}$
 and every tuple $(a_1,\ldots,a_{\ar(R_i)}) \in R_i^\bfA$,
 it holds that
 the tuple $(h(a_1),\ldots,h(a_{\ar(R_i)}))$ belongs to $R_i^\bfB$.
The relational structure
$\bfA[\phi]$
corresponding to a conjunctive query $\phi$
(where all quantifiers are in front)
has as domain the set of variables that occur,
and, interprets each symbol $R$ as the relation
containing each tuple that $R$ is applied to, in $\phi$;
for example, $\bfA[\psi]$ for $\psi = \exists v_1 \exists v_2 \exists v_3
(E(v_1,v_2) \wedge E(v_2, v_3) \wedge E(v_3, v_1))$ is equal to
$(\{ v_1, v_2, v_3 \}; E^{\bfA[\psi]})$,
where $E^{\bfA[\psi]} = \{ (v_1, v_2), (v_2, v_3), (v_3, v_1) \}$.

The problem of conjunctive query evaluation
and the relational homomorphism problem
are, in fact, both considered
formulations of the \emph{constraint satisfaction problem}~\cite{FederVardi99-structure}.
The complexity-theoretic aspects of these problems
have been heavily studied over the years;
see for example~\cite{AbiteboulHullVianu95-foundationsdatabases,PapadimitriouYannakakis99-database,GottlobLeoneScarcello01-complexityacyclic,GottlobLeoneScarcello02-hypertreedecomposisions,Grohe07-otherside,SchweikardtSchwentickSegoufin09-querylanguages,Marx10-tractablehypergraph,BartoKozik09-boundedwidth,ChenMueller17-hierarchy,ChenMueller15-fineclass}
and the references therein.

\subsection{Contributions} \label{subsect:contributions}

Despite the centrality of query evaluation as a
computational problem in the field of databases,
and the ability of the property testing framework
to provide meaningful theoretical results
for problems where examining the entire input is
exceedingly costly,
there has not been---to the best of our knowledge---any
systematic study of database query evaluation
from the viewpoint of property testing, on general relational
databases.
In this article, we initiate such a study:
we
present a natural oracle model under which
relational structures can be accessed; and,
we establish a dichotomy theorem that, in a certain precise sense,
completely classifies the conjunctive queries
that give rise to properties that are constant-query testable.

For cleanliness and ease of discussion,
we work primarily with the homomorphism formulation of
conjunctive query evaluation.
When $\bfA$ and $\bfB$ are structures having the same vocabulary,
let us say that $\bfB$ is \emph{$\bfA$-free} if
$\bfB$ does not admit a homomorphism from $\bfA$.
We characterize precisely when,
for a structure $\bfA$,
one has constant-query testability of
the property of $\bfA$-freeness;
this property is here
also referred to as
the property of \emph{homomorphism inadmissibility from $\bfA$}.
Stated in the parlance of conjunctive queries, we characterize precisely
when, for a conjunctive query $\phi$,
one has constant-query testability of
the property of being false under $\phi$;
as each problem studied corresponds to a fixed conjunctive query $\phi$
and takes as input a structure $\bfB$ to be evaluated on $\phi$,
our study can be taken as a property testing characterization
of the data complexity of each conjunctive query.

{\bf Model studied.}
Let $\bfA$ and $\bfB$ be relational structures over the same vocabulary,
and let $\epsilon \in (0,1)$;
the structure $\bfB$
is called \emph{$\epsilon$-far} from $\bfA$-free if it is necessary to remove more than an $\epsilon$-fraction of tuples
from $\bfB$ in order to ensure that
it does not admit a homomorphism from $\bfA$;
it is called \emph{$\epsilon$-close} to $\bfA$-free otherwise.
In the problem of testing homomorphism inadmissibility from $\bfA$,
an algorithm is given an error parameter $\epsilon \in (0,1)$
and is permitted
query access to a
relational structure $\bfB$ with the same vocabulary as $\bfA$;
its goal is to
 distinguish the case that $\bfB$ is
 $\bfA$-free from the case that $\bfB$ is $\epsilon$-far from $\bfA$-free.
An algorithm achieving this goal
is called a \emph{tester for $\bfA$-freeness}.
Such a tester is \emph{constant-query}
if the number of queries to $\bfB$ is a function only of
$\epsilon$;
is said to \emph{use $Q$ queries},
where $Q$ is a family of functions on the natural numbers,
if for each $\epsilon$ there exists $q_\epsilon \in Q$ such that
at most $q_{\epsilon}(|B|)$ queries are used when the presented
error parameter is $\epsilon$;
and, is of \emph{one-sided error} if it
 accepts every structure $\bfB$ that is $\bfA$-free.

In our model, access to the structure $\bfB$ is via queries
to
what we term the \emph{completion oracle}.
Consider a relation $R^{\bfB}$ of arity $r$ in $\bfB$
and a tuple $\bfq = (q_1,\ldots,q_r)\in {(B \cup \set{*})}^r$,
where $B$ is the domain of $\bfB$ and $*$ is a special wildcard symbol.
A tuple $\bft = (t_1,\ldots,t_r)$ is called a \emph{completion}
of $\bfq$ in $R^{\bfB}$
if
(i) $t_i = q_i$ for every $i \in \set{1,\ldots,r}$ with $q_i \in B$
and
(ii) $\bft \in R^{\bfB}$.
The completion oracle
allows two types of queries; for each, the oracle
receives a relation symbol $R$
and a tuple
$\bfq = (q_1,\ldots,q_r)\in {(B \cup \set{*})}^r$:
\begin{itemize}
\item To a \emph{completion query}, the oracle returns
a random completion of $\bfq$ in $R^{\bfB}$;
if no such completion exists, it returns a special symbol $\bot$.
\item To a \emph{size query}, the oracle returns
the number of completions of $\bfq$ in $R^{\bfB}$.
\end{itemize}
We  discuss the relation between
the completion oracle and the established \emph{general graph} model
later,
in Section~\ref{subsect:model-relationship}.

One may wonder why we do not test homomorphism admissibility instead of homomorphism inadmissibility.
The reason is that,
for any fixed structure $\bfA$,
 we can add a constant number of tuples to a structure $\bfB$
to realize a homomorphism from $\bfA$ to $\bfB$.
Hence, we need to choose $\epsilon = O(1/\|\bfB\|)$ to make the testing problem meaningful, where $\|\bfB\|$ is the total number of tuples in $\bfB$; otherwise, no $\epsilon$-far instance exists and we can always answer that there is a homomorphism from $\bfA$ to $\bfB$.
However, when $\epsilon$ is so small, we can obtain the whole structure $\bfB$ with $O(1/\epsilon)$ queries, and can then exactly check whether there is no homomorphism from $\bfA$ to $\bfB$.
Hence, testing homomorphism admissibility is an uninteresting problem,
under the employed lens of property testing.

We remark that the introduced model may permit
the study of further properties,
beyond $\bfA$-freeness,
that are \emph{anti-monotone};
here, by an \emph{anti-monotone} property, we mean a property
on structures such that,
when $\bfB$ is a structure that does not satisfy the property,
adding tuples to $\bfB$ always results in a structure
that also does not satisfy the property.
We leave such a study as a potential avenue for future work.

{\bf Main results.}
The primary technical contribution of this work is
a dichotomy theorem that completely characterizes
the structures $\bfA$ for which
the property of $\bfA$-freeness
is constant-query testable with one-sided error.
In the following,
$n$ denotes the domain size $|B|$ of the structure $\bfB$ being tested.

\begin{theorem}\label{the:main}
  Let $\bfA$ be any relational structure.
  The property of $\bfA$-freeness is constant-query testable with one-sided error if and only if the core of $\bfA$ is $\alpha$-acyclic.
  In particular, when the core of $\bfA$ is not $\alpha$-acyclic,
  there exists a constant $c \geq 1$ such that
  testing $\bfA$-freeness, with one-sided error, requires $\Omega(n^{1/c})$ queries.
\end{theorem}
Here, \emph{$\alpha$-acyclicity} is an established notion of acyclicity
 for hypergraphs,
 and is applied to structures in a natural way; precise definitions are given in
Section~\ref{sec:pre}, where
the definition of  \emph{core} is also given.
Indeed, the notion of \emph{$\alpha$-acyclicity} is classical in database theory;
it is known that there is a polynomial-time algorithm
for the problem of deciding if an $\alpha$-acyclic structure admits
a homomorphism to another structure
(see for example~\cite[Chapter 6.4]{AbiteboulHullVianu95-foundationsdatabases}).

The just-presented dichotomy theorem may be restated in database parlance, as follows.
By a \emph{positive first-order query},
we here refer to a first-order logical sentence that
does not contain negation; such a sentence evaluates
to true or false when given a structure over the query's vocabulary.
When $\phi$ is a positive first-order query, we say that a structure $\bfB$ is
\emph{$\epsilon$-far} from $\phi$-falsity if it is necessary to remove
more than an $\epsilon$-fraction of tuples from $\bfB$ to ensure
that $\phi$ evaluates to false on $\bfB$;
we define the problem of testing $\phi$-falsity analogously to
how we defined the problem of testing $\bfA$-freeness.

\begin{theorem} (Restatement of Theorem~\ref{the:main}'s dichotomy)
Let $\phi$ be any conjunctive query; the property of $\phi$-falsity
is constant-query testable with one-sided error if and only if
the core of $\bfA[\phi]$ is $\alpha$-acyclic.
\end{theorem}

In the area of databases,
a heavily studied generalization of the class of conjunctive queries
is the class of so-called 
\emph{unions of conjunctive queries}~\cite{Chen14-existentialpositive,Fagin:2005kn}.
We formalize these logically as follows.
A \emph{union of conjunctive queries} is a disjunction of conjunctive queries;
so, the class of structures on which it evaluates to true
is equal to the union, over its constituent conjunctive queries $\theta$,
of the structures on which $\theta$ evaluates to true.

From Theorem~\ref{the:main}, we may derive a more general dichotomy theorem,
which describes the testability of each union of conjunctive queries.
It is known that any union $\phi$ of conjunctive queries
can be computably rewritten to a form,
which we call the \emph{normalization of $\phi$},
wherein none of its disjuncts are redundant;
see~\cite[Section 3]{BovaChen14-width-ep}.\footnote{In particular, a union of conjunctive queries as defined here is,
	in~\cite{BovaChen14-width-ep}, referred to
	as an existential positive sentence in disjunctive form;
	by the \emph{normalization of $\phi$}, we mean
	an existential positive sentence in disjunctive form
	that is \emph{irredundant} in the sense of~\cite{BovaChen14-width-ep}.
}

\begin{corollary}\label{cor:union-of-cqs}
Let $\phi$ be any union of conjunctive queries,
and let $\phi' = \phi_1 \vee \cdots \vee \phi_k$ be
the normalization of $\phi$.
The property of $\phi$-falsity
is constant-query testable with one-sided error if and only if
the following condition holds:
for each $i \in [k]$,
the core of $\bfA[\phi_i]$ is $\alpha$-acyclic.
When this condition does not hold,
there exists a constant $c \geq 1$ such that
  testing $\phi$-falsity, with one-sided error, requires $\Omega(n^{1/c})$ queries.
\end{corollary}

In this article,
we also consider the injective version of testing $\bfA$-freeness~\cite{embedding}.
We say that a relational structure $\bfB$ is \emph{injectively $\bfA$-free} if there is no injective homomorphism from $\bfA$ to $\bfB$.
Then, the testing problem is to
 distinguish the case that $\bfB$ is injectively $\bfA$-free from the case that $\bfB$ is $\epsilon$-far from being so.
By building on and slightly modifying the positive result of Theorem~\ref{the:main},
we show the positive testability result that,
for any relational structure $\bfA$ that is $\alpha$-acyclic,
injective $\bfA$-freeness is constant-query testable
\ifthenelse{\boolean{fullversion}}{(Theorem~\ref{the:alpha-acyclic-injective})}{}.

{\bf Oracle implementation issues.}
In our view, the theory developed in this article evidences that
our oracle model is mathematically robust and natural; again, we regard
the presentation of this model as a contribution of this article.
In Theorem~\ref{the:main} and the other dichotomy theorems presented above,
however,
oracle queries are treated as primitives; these theorems are agnostic
about the particulars of how oracle queries are answered.
While we hope that this article will engender further interaction
between property testing and database theory,
we here would like to point out that the oracle queries of our model
can be resolved in polylogarithmic time in the case that the
relations of the structure $\bfB$ being tested are represented as range trees~\cite{Bentley:1979cv}, where we identify elements in $B$ with values $1,\ldots,|B|$ through an arbitrary one-to-one correspondence.
Precisely, consider
the representation of each relation $R^{\bfB}$ of $\bfB$
by a range tree $T_{R^\bfB}$ of dimension $\ar(R)$.
In each one-dimensional range tree $T$ used in $T_{R^\bfB}$, a leaf node with value $b \in B$
is annotated with the number of
tuples, in the relation represented by $T$, that begin with $b$;
and each internal node $v$ is annotated with the sum of the annotations
over all leaf descendants of $v$.
It is straightforward to verify that both completion queries
and size queries concerning a relation $R^{\bfB}$ can be answered in
time $O(\log^d |R^{\bfB}|)$, where $d = \ar(R)$.

\subsection{Relationship between our model and the general graph model}\label{subsect:model-relationship}

We may view each undirected graph as a relational structure
$\bfG = (V; E^{\bfG})$ where $E^{\bfG}$ is a symmetric binary relation.
Previous work~\cite{Alon:2008gn,Kaufman:2004vg}
studied the problem of testing, relative to a fixed graph $\bfH$,
 whether a given graph $\bfG$ is
 injectively $\bfH$-free or is $\epsilon$-far from
 being injectively $\bfH$-free.\footnote{Let us remark that, in the previous literature,
 \emph{injective $\bfH$-freeness} as defined in this article
 is sometimes referred to simply
 as \emph{$\bfH$-freeness}.
 In the present article,
 as the notion of homomorphism is of primary interest,
 we adhere to the terminology established in
 Section~\ref{subsect:contributions}.
 Note that for a graph $\bfG$ to be injectively
 $\bfH$-free is equivalent to $\bfG$ not containing $\bfH$
 as a subgraph (up to isomorphism).
 }
Here, one often assumes
that each graph $\bfG = (V; E^{\bfG})$ to be tested has
$|E^{\bfG}| = \Omega(|V|)$, to avoid technical triviality.
In the \emph{general graph model}~\cite{Alon:2008gn,Kaufman:2004vg},
in addition to $|V|$ and $|E^\bfG|$, the graph $\bfG$ is provided through query access;
the following four types of oracles are provided:
\begin{itemize}
  \itemsep=0pt

\item Random edge oracle: It returns a random edge chosen uniformly at random from $E$.

\item Random neighbor oracle: Given a vertex $u \in V$, it returns a random neighbor of $v$ if a neighbor exists, and a special symbol $\bot$ otherwise.

\item Adjacency oracle: Given a vertex pair $(u,v) \in V^2$, it returns yes if the edge $\set{u,v}$ exists in $E$, and no otherwise.

\item Degree oracle: Given a vertex $v \in V$, it returns the degree of $u$.
\end{itemize}

Allowing these four types of oracles is equivalent to
allowing queries to the completion oracle; this is seen as follows:
\begin{itemize}
  \itemsep=0pt
\item A query to the random edge oracle corresponds to
the completion query $(E, (*,*))$ to the completion oracle.

\item The query $u \in V$ to the random neighbor oracle
corresponds to the two completion queries $(E, (u,*))$
and $(E, (*,u))$ to the completion oracle.
A response of $v$ from the random neighbor oracle
corresponds to the response of $(u,v)$ or $(v,u)$
from the completion oracle (respectively);
a response of $\bot$ from the random neighbor oracle
corresponds to the response of $\bot$ from the completion oracle.

\item The query $(u,v) \in V^2$ to the adjacency oracle
corresponds to the completion query $(E, (u,v))$ to the completion oracle.
The response of \emph{yes} from the adjacency oracle
corresponds to the response of $(u,v)$ from the completion oracle;
the response of \emph{no} from the adjacency oracle
corresponds to the response of $\bot$ from the completion oracle.
\item The query $u \in V$ to the degree oracle
corresponds to the size queries $(E, (u,*))$
and $(E, (*,u))$ to the completion oracle.
\end{itemize}
We note that the answer to the size query $(E,(*,*))$ to the completion oracle is $|E^\bfG|$, and that the answer to the size query $(E,(u,v))$ to the completion oracle can be answered by asking
the adjacency oracle whether $u$ and $v$ are adjacent:
the answer is $1$ if $u$ and $v$ are adjacent and $0$ otherwise.

It was previously shown
that the property of injective $\bfH$-freeness is constant-query testable
if $\bfH$ is a path of a constant length~\cite{Iwama:2018bs}, and requires $\omega(|V|^{1/2})$ queries when $\bfH$ is non-bipartite~\cite{Alon:2008gn}.

We can conceive of a generalization of the general graph model for $k$-uniform hypergraphs where, by specifying a set $S$ of vertices of size at most $k$,
an algorithm can obtain
a random hyperedge containing $S$,
as well as the number of such hyperedges.
Our model encompasses testing injective $\bfH$-freeness in such a model,
though to the best of our knowledge, no prior work focused on this problem.
Although testing properties of hypergraphs has been extensively studied~\cite{Fox:2016bp,Joos:2017tb,Rodl:2007dt},
most works considered the \emph{dense model},
in which we can ask whether there is a hyperedge spanning specific $k$ vertices.

\subsection{Proof Idea}

We explain the proof idea of Theorem~\ref{the:main}.
We first note that, for a structure $\bfA$ and its core $\bfC$, testing $\bfA$-freeness and $\bfC$-freeness are equivalent.
Hence, we can assume that $\bfA$ itself is a core.

{\bf Algorithm.}
We here explain the idea of our algorithm, which only uses completion queries
(that is, it does not use size queries).
By the definition of $\alpha$-acyclicity, we can order tuples in relations of $\bfA$ as $\bfa^1,\ldots,\bfa^m$ so that,
for each $i > 1$,
all the variables in $\bfa^i$ that appeared in $\bfa^1,\ldots,\bfa^{i-1}$ are contained in a single tuple $\bfa^j$ with $j < i$.
Let $\set{\bfa^k}$ be the set of variables that appear in $\bfa^k$.
Then, starting with an empty map $f^0:\emptyset \to B$, we iteratively create a partial map $f^k : \bigcup_{\ell \in \set{1,\ldots,k}}\set{\bfa^\ell} \to B$ from $f^{k-1}$ by assigning values to unassigned variables in $\bfa^k$ using completion queries.
If a completion query returns $\bot$ along the way, then
the algorithm accepts,
claiming that there is no homomorphism from $\bfA$ to $\bfB$.
Otherwise, the algorithm rejects,
claiming that $\bfB$ is $\epsilon$-far from $\bfA$-freeness.

If the algorithm successfully obtains a map $f^m$
without ever getting $\bot$ from the completion oracle,
this map always gives rise to a homomorphism from $\bfA$ to $\bfB$.
This means that the algorithm always accepts when $\bfB$ is $\bfA$-free because there is no homomorphism from $\bfA$ to $\bfB$.

The proof that the algorithm,
with high probability,
rejects any structure $\bfB$ that is $\epsilon$-far from $\bfA$-free
consists of the following three steps.
First, we take a large set of homomorphic images of $\bfA$ in $\bfB$ that are disjoint in a certain sense, and construct a substructure $\bfB'$ of $\bfB$ consisting of tuples used in the homomorphic images.
Then, we show that, if the probability that the algorithm rejects $\bfB'$ is high, then the probability that the algorithm rejects $\bfB$ is also high.
Finally, we show that the probability that the algorithm rejects $\bfB'$ is high using the disjointness property, which establishes the correctness of the algorithm.

{\bf Lower bound.}
We now explain how we obtain a super-constant lower bound for testing $\bfA$-freeness when it is not $\alpha$-acyclic.
We first show that there exists a structure $\bfA'$ such that (i) a super-constant lower bound for testing $\bfA'$-freeness implies a super-constant lower bound for testing $\bfA$-freeness, (ii) $\bfA'$ is symbol-distinct in the sense that each relation contains exactly one tuple, and (iii) the hypergraph constructed by regarding each tuple in $\bfA'$ a hyperedge is (essentially) either a cycle or a
\emph{$k$-simplex},
which is, up to isomorphism, a $(k+1)$-vertex hypergraph with a hyperedge $\set{1,\ldots,k+1}\setminus \set{i}$ for each $i \in \set{1,\ldots,k+1}$.
From (i), it suffices to show a super-constant lower bound for $\bfA'$.
We consider the cycle and simplex cases separately.

When the hypergraph is a cycle, we can show a super-constant lower bound by using the known lower bound for testing triangle-freeness in the general graph model~\cite{Alon:2008gn} by a relatively simple reduction.

When the hypergraph is a simplex, we can show a super-constant lower bound by generalizing the argument used in~\cite{Alon:2008gn}.
Namely, we consider a distribution $\caD$ over $k$-uniform hypergraphs generated as follows: We partition the vertex set into $k+1$ parts $V_1,\ldots,V_{k+1}$ of the same size, and for each set of $k$ parts $\set{V_{i_1},\ldots,V_{i_k}}$, we take random $k$-dimensional matchings spanning them.
We can show that the created hypergraph $G$ has many disjoint $k$-simplices and hence is $\Omega(1)$-far from $k$-simplex-freeness with high probability.
Note that a one-sided error tester $\caA$ can reject a hypergraph only when it finds a $k$-simplex as a subgraph.
Considering when $\caA$ finds the last hyperedge of a $k$-simplex, this means that $\caA$ should query a tuple $\bfq$ with wildcard symbols and the returned completion $\bft$ of $\bfq$ should consist only of vertices that already appeared in the process of $\caA$.
From the random construction of the hypergraph, however, we can show that the wildcard symbols in $\bfq$ are completed by vertices that have never appeared in the process as long as the number of queries is constant.

\subsection{Related work}
To the best of our knowledge, testing homomorphism inadmissibility,
as formalized in this article,
 has not been systematically studied in the literature.
The most relevant line of research is testing subgraph-freeness in the general graph model, as discussed above.
There are several variations on the general graph model regarding the provided oracles.
In~\cite{Goldreich:2008in,Kaufman:2004vg}, a model with (a non-random version of) the random neighbor, adjacency, and degree oracles is considered.
In~\cite{Dasgupta:2014kg}, a model with essentially the random edge oracle is considered.
In~\cite{Iwama:2018bs}, a model with all of the random edge, random neighbor, adjacency, and degree oracles is considered.
From the perspective of relational structures, it is natural to consider the model with all the oracles.

The computational problem of deciding, given a pair $(\bfA,\bfB)$ of relational structures
over the same vocabulary, if there exists a homomorphism from $\bfA$ to $\bfB$,
is a formulation of the constraint satisfaction problem, which is well-known to be NP-complete.
There is an extensive literature that aims to identify restricted cases of this problem
that are polynomial-time tractable.  One main stream in this literature studies
restricted cases where the target structure $\bfB$ is fixed, see for example~\cite{BartoKozik09-boundedwidth,ChenLarose17-metaquestions,Bulatov17-dichotomy,Zhuk17-dichotomy} and the references therein.
Another stream, which is more closely aligned with the present study, considers
cases that are based on limiting the permitted source structures $\bfA$,
see for example~\cite{GottlobLeoneScarcello01-complexityacyclic,GottlobLeoneScarcello02-hypertreedecomposisions,Grohe07-otherside,Chen14-existentialpositive,ChenMueller15-fineclass,ChenMueller17-hierarchy}
and the references therein.

In property testing, constraint satisfaction problems
have been extensively studied;
previous work can be classified into two categories:
\begin{itemize}
\item The first category is testing whether a given CSP instance, provided through a query access, is satisfiable.
In the \emph{dense model}, one restricts attention to CSP instances
whose constraints have a fixed arity $k$, and  an instance is considered $\epsilon$-far from being satisfiable if one needs to remove more than $\epsilon n^k$ constraints in order to make it satisfiable, where $n$ is the number of variables.
In this model, any CSP is known to be constant-query testable~\cite{Alon:2003jp,Alon:2002ua,Sohler:2012kqa}.
In the \emph{bounded-degree model}, one restricts attention to CSP instances where each variable participates in a constant number, say, $d$, of constraints, and an instance is considered $\epsilon$-far from being satisfiable if one needs to remove more than $\epsilon d n$ constraints in order to make it satisfiable.
In this model, a characterization is obtained via
the LP integrality gap of the CSP~\cite{Yoshida:2011da}.

\item The second category is testing whether a given assignment to a CSP instance, through query access, is a satisfying assignment.
This testing problem has been considered relative to fixed target structures $\bfB$.
Here, the CSP instance is explicitly provided to the algorithms.
It was shown that $2$-SAT is testable with $O(\sqrt{n})$ queries~\cite{Fischer:2002ev} and testing $3$-SAT requires $\Omega(n)$ queries~\cite{BenSasson:2005we}, where $n$ is the number of variables.
In~\cite{Yoshida:2014zn}, universal algebra is used to systematically study this problem.
A characterization of constant-query testable Boolean CSPs was achieved in~\cite{Bhattacharyya:2013fa}, and it was generalized to finite-domain CSPs in~\cite{Chen:2016wf}.
\end{itemize}

We also mention that, in the context of database theory,
Ben-Moshe et al.~\cite[Section 6]{BenMoshe11}
studied the testability of a notion called \emph{near-sortedness}.
In their study, near-sortedness is a property of relations presented as
an ordered list of tuples,
whereas (as mentioned) in the present work, the oracle access to the structure
being tested is not tied to a fixed representation of relations.
In~\cite{Adler:2018jv}, the authors introduced a bounded-degree model for databases and showed that, if the databases have bounded tree-width, then every query definable in monadic second-order logic with modulo counting is testable with a constant number of oracle queries.

\subsection{Organization}

We introduce notions used throughout this paper in Section~\ref{sec:pre}.
We show that $\bfA$-freeness is constant-query testable if the core of $\bfA$ is $\alpha$-acyclic in Section~\ref{sec:upper-bound}; and, that it is not constant-query testable otherwise, in Section~\ref{sec:lower-bound}.
We provide the proof of Corollary~\ref{cor:union-of-cqs}, our result on unions of conjunctive queries, in Section~\ref{sect:ucqs}.

\section{Preliminaries}\label{sec:pre}
For each integer $n$, we use $[n]$ to denote the set $\set{1,2,\ldots,n}$.
We use bold symbols, such as $\bft$, to denote tuples.
For a tuple $\bft=(t_1,\ldots,t_k)$,
we use $\set{\bft}$ to denote the set $\set{t_1,\ldots,t_k}$.
For a function $f:A \to B$
and a tuple $\bfa=(a_1,\ldots,a_k)$ of elements of $A$, we define $f(\bfa)$ to be the tuple $(f(a_1),\ldots,f(a_k))$.
We use double brackets, as in $\multiset{a,a,b,c}$, to denote a multiset.
When $S$ is a multiset, we sometimes use $\# S$ to denote its size.

\subsection{Relational structures and homomorphisms}
A \emph{vocabulary} is a set of relational symbols, where each
relational symbol $R$ has an associated \emph{arity},
which is a positive integer and is
denoted by $\ar(R)$.
Throughout this paper, we assume that each vocabulary is finite.
A \emph{relational structure}
(for short, \emph{structure})
with vocabulary $\sigma$ is
a pair of the form $\bfA = (A, \set{R^\bfA}_{R \in \sigma})$,
where $A$ is a set, called the \emph{domain} of $\bfA$, and each
relation $R^{\bf A}$ %
is a subset of $A^{\ar(R)}$.
We define a \emph{multiset relational structure}
(for short, \emph{multiset structure})
similarly, with the difference that each relation
$R^{\bf A}$ is a multiset whose elements are drawn from $A^{\ar(R)}$.
In this article, each relational structure
and each multiset relational structure under discussion
is assumed to be \emph{finite} in the sense that its domain is finite.
We generally use  the letters
$\bfA,\bfB,\bfC,\ldots$ to denote structures and multiset structures,
and the letters
$A,B,C,\ldots$ to denote their respective domains.
For a structure or multiset structure
$\bfA=(A,\set{R^\bfA}_{R\in \sigma})$, we denote by $\|\bfA\|$ the total number of tuples in the relations of $\bf A$, that is, $\sum_{R \in \sigma}|R^\bfA|$.

For a structure $\bfA = (A,\set{R^\bfA}_{R \in \sigma})$
and a structure or multiset structure
$\bfB = (B,\set{R^\bfB}_{R \in \sigma})$
having the same vocabulary
$\sigma$, a \emph{homomorphism} from $\bfA$ to $\bfB$ is a mapping $h : A \to B$ such that, for any $R \in \sigma$ and $\bft = (a_1,\ldots,a_r)\in R^\bfA$, the image $h(\bft)=(h(a_1),\ldots,h(a_r))$ belongs to $R^\bfB$.

An \emph{endomorphism}
of a structure
$\bfC = (C,\set{R^\bfC}_{R \in \sigma})$
is a homomorphism from $\bfC$ to itself.
A \emph{core} is a structure such that each endomorphism thereof
is a bijection.
A \emph{substructure}
of a structure
$\bfA = (A,\set{R^\bfA}_{R \in \sigma})$
is a structure
$\bfC = (C,\set{R^\bfC}_{R \in \sigma})$
such that $C \subseteq A$ and $R^{\bfC} \subseteq R^{\bfA}$
for each $R \in \sigma$.\footnote{Since a substructure $\bfC$ of a structure $\bfA$ is, by definition,  a structure itself, it indeed holds that
$R^{\bfC} \subseteq R^{\bfA} \cap C^{\ar(R)}$,
for each symbol $R$.}
A \emph{core of a structure
$\bfA = (A,\set{R^\bfA}_{R \in \sigma})$}
is a substructure of $\bfA$
that is a core and
that receives a homomorphism from $\bfA$.
It is known that each structure has at least one core,
and that all cores of a structure are isomorphic~\cite{HellNesetril92-core};
following a standard abuse, we thus speak of \emph{the} core
of a structure.

\begin{example}
Undirected graphs can be naturally represented as structures having a single binary relation that is symmetric.
Under this representation,
the following are well-known and straightforward to verify:
\begin{itemize}
\item  any odd cycle is a core (that is, each endomorphism of an odd cycle is a bijection),
and hence the core of any odd cycle is the odd cycle itself;
and,

\item the core of any even cycle is a single edge,
that is, a two-vertex graph having a single edge between
the two vertices.  Indeed, the core of any bipartite graph (which contains at least one edge)
is a single edge.

\end{itemize}
\end{example}

\subsection{Hypergraphs and $\alpha$-acyclicity}
Let $V$ be a set, and let $S, S_1, \ldots, S_k \subseteq V$.
We say that $S$ is an \emph{ear} with respect to
$S_1, \ldots, S_k$ when
either $k = 0$ or
there exists $j \in [k]$ such that
$S \cap (\bigcup_{i \in [k]} S_i) \subseteq S_j$.
A hypergraph $G=(V,\caE)$ with $m$ hyperedges is
called \emph{$\alpha$-acyclic} if there exists an ordering
$S_1, \ldots, S_m$
of the hyperedges so that, for each $i \in [m]$, the hyperedge $S_i$ is an ear with respect to the previous edges $S_1,\ldots,S_{i-1}$.
For a relational structure $\bfA = (A,\set{R^\bfA}_{R \in \sigma})$, its associated hypergraph is defined as $G_\bfA=(A,\caE)$, where
$\caE = \set{\set{\bft} \mid  R \in \sigma, \bft \in R^\bfA}$.
We say that a relational structure $\bfA$ is \emph{$\alpha$-acyclic}
if its associated hypergraph $G_\bfA$ is $\alpha$-acyclic.

We say that a hypergraph $G=(V,\caE)$ on $k+1$ vertices is \emph{($k$-)simplex} if,
up to isomorphism,
$V = [k+1]$ and
 $\caE$ consists of the $k+1$ hyperedges of the form
 $[k+1] \setminus \set{i}$, where $i \in [k+1]$.
We say that a hypergraph $(V,\caE)$ is a \emph{cycle} if,
up to isomorphism,
it holds that $V = \{ v_1, \ldots, v_m \}$ and
$\caE = \set{ (v_i,v_{i+1}) \mid i \in [m]}$, where $v_{m+1} = v_1$.

Let $G = (V,\caE)$ be a hypergraph.
For a subset $S \subseteq V$,
the \emph{hypergraph induced by $S$}, denoted by $G[S]$, is defined as
$(S, \set{ e \cap S \mid e \in \caE } \setminus \{ \emptyset \})$.
The hypergraph $\caM(G)$ is defined as
$(V, \set{ e \in \caE \mid \not\exists f \in \caE, e \subsetneq f })$,
that is, it is the hypergraph whose hyperedges are the inclusion maximal
hyperedges of $G$.
We will make use of the following known characterization;
we recommend~\cite[Characterization 2.5]{BB16}  as a reference.

\begin{theorem}\label{the:characterization-of-alpha-acyclic}
A hypergraph $G = (V,\caE)$ is not $\alpha$-acyclic
if and only if there exists $S \subseteq V$ with $|S| \geq 3$
such that $\caM(G[S])$ is either a simplex or a cycle.
\end{theorem}

\subsection{Testing Homomorphism Inadmissibility}

\emph{Throughout this paper, whenever we speak of testing
$\bfA$-freeness,
$\bfA$ will be a structure, and the object $\bfB$ being tested
will be a multiset structure.}
The study of relational databases whose relations are multisets is
established and natural in the database context; it is in line with
so-called \emph{multiset} or \emph{bag} semantics~\cite{ChaudhuriV93}.

For a  structure $\bfA$, we say that a multiset structure $\bfB$ is
\emph{$\epsilon$-far from $\bfA$-freeness}
if we must remove more than $\epsilon \|\bfB\|$ tuples from relations in $\bfB$ in order to ensure that it does not receive a homomorphism from $\bfA$.
Formally, a multiset structure $\bfB$ is
\emph{$\epsilon$-far from $\bfA$-freeness}
if, for any structure $\bfB'$ obtained from $\bfB$ by removing tuples
from relations, $\| \bfB \| - \| \bfB' \| \leq \epsilon \|\bfB \|$
implies that $\bfB'$ admits a homomorphism from $\bfA$.

When testing a multiset structure $\bfB$, two types of queries
are permitted.  Each type of query receives
a pair $(R,\bfq)$ of a relation symbol $R$ in the vocabulary of $\bfB$ and a tuple $\bfq \in {(B \cup \set{*})}^{\ar(R)}$, where $* \not \in B$ is a special symbol.
Let $S_\bfB(t) = B_1 \times \cdots \times B_{\ar(R)}$, where $B_i = B$ if $t_i = *$ and $B_i = \set{t_i}$ otherwise.
We call each tuple in $S_\bfB(\bfq)$ a \emph{completion} of $\bfq$.
The answer to a
\emph{completion query} is a random completion of $\bfq$ in $R^\bfB$,
if a completion exists;
formally, the result is a tuple chosen uniformly at random
from the multiset $\multiset{ \bft \in R^\bfB \mid \bft \in  S_\bfB(\bfq) }$
when this multiset is non-empty, and a special symbol $\bot$ otherwise.
The answer to a
\emph{size query} is the size of the multiset
$\multiset{ \bft \in R^\bfB \mid \bft \in  S_\bfB(\bfq) }$.
We use $\deg_{R,\bfB}(\bfq)$ to denote
$|\multiset{ \bft \in R^\bfB \mid \bft \in S_\bfB(\bfq)}|$, which is the number of completions of $\bfq$ in $R^\bfB$.

For a structure $\bfA$ with a finite vocabulary $\sigma$, we say that an algorithm is a \emph{tester} for $\bfA$-freeness if,
given a domain $B$, query access to a multiset structure $\bfB$ with the vocabulary $\sigma$, and an error parameter $\epsilon \in (0,1)$, it accepts with probability at least $2/3$ if $\bfB$ is $\bfA$-free, and it rejects with probability at least $2/3$ if $\bfB$ is $\epsilon$-far from $\bfA$-freeness.
The notions of injective $\bfA$-freeness, of a
tester for injective $\bfA$-freeness, etc.
are defined analogously using injective homomorphisms.

\section{Constant-Query Testers for \texorpdfstring{$\alpha$}{a}-Acyclic Relational Structures}\label{sec:upper-bound}

In this section, we present a constant-query tester
for
homomorphism inadmissibility from an $\alpha$-acyclic relational structure.
This positive result applies directly to any $\alpha$-acyclic relational structure,
even those that are not cores.
We describe our algorithm in Section~\ref{subsec:algorithm} and analyze it in Section~\ref{subsec:analysis}.
The modification needed to test injective homomorphism inadmissibility is given in Section~\ref{subsec:injective}.
Throughout the section,
we fix $\bfA$ to be an $\alpha$-acyclic relational structure,
and let $m = \|\bfA\|$.

\subsection{Algorithm}\label{subsec:algorithm}

First, the algorithm preprocesses $\bfA$ before querying $\bfB$.
Consider the set
$\caA = \set{(\bfa, R) \mid R \in \sigma, \bfa \in R^\bfA}$.
By the $\alpha$-acyclicity of $\bfA$, there is an ordering
$(\bfa^1,R_1),\ldots,(\bfa^m,R_m)$ of the elements in $\caA$ such that,
for each $k \in [m]$, the set
$\set{\bfa^k}$ is an ear with respect to the sets
$\set{\bfa^1},\ldots,\set{\bfa^{k-1}}$.

Our algorithm is a random process,
called \Call{HI-Test$_\bfA$}{} and detailed precisely in
Algorithm~\ref{alg:alpha-cyclic}.
We give an overview here.
The algorithm \Call{HI-Test$_\bfA$}{} proceeds in $m$ stages.
Starting with an empty map $f^0: \emptyset \to B$,
the algorithm attempts to
construct a partial map $f^k$: $\bigcup_{\ell \in [k]} \set{\bfa^\ell} \to B$ in the $k$-th stage.
More specifically, in the $k$-th stage,
the algorithm tries to extend the
domain of $f^{k-1}$ so that it is defined on
all entries of the tuple $\bfa^k$, by making a completion query;
if the answer to the query is $\bot$, then the algorithm gives up
and claims that there is no homomorphism, by accepting.
Let us say that the algorithm is \emph{$k$-successful} if it did not fail in the first $k$-th stages and succeeded to define $f^k$.
If the algorithm is $m$-successful, then the algorithm
claims that there is a homomorphism, by rejecting.

\begin{algorithm}[t!]
  \caption{}\label{alg:alpha-cyclic}
  \begin{algorithmic}[1]
    \Procedure{HI-Test$_{\bfA}$}{$B$, $\bfB$}\\
      \Comment{$\bfB$ is a multiset structure with domain $B$ and having the same vocabulary as $\bfA$.}
      \State Let $(\bfa^1,R_1),\ldots,(\bfa^m,R_m)$ be the ordering, as explained in the text.
      \State Let $f^0: \emptyset \to B$ be the empty map.
      \For{$k=1$ to $m$}
        \State Let $\bfq^k = (q^k_1,\ldots,q^k_{\ar(R_k)})$ be the tuple defined as
        \[
          q^k_i = \begin{cases}
            f^{k-1}(a^k_i) & \text{if }a^k_i \in \cup_{\ell \in [k-1]} \set{\bfa^\ell}\\
            * & \text{otherwise.}
          \end{cases}
        \]
        \State Pose the completion query $(R_i, \bfq^k)$ to $\bfB$.
        \If{The answer is $\bot$}
          \State Accept.
        \Else
          \State Let $\bfb^k = (b^k_1,\ldots,b^k_{\ar(R_k)})$ be the answer.
          \State Define $f^k: \bigcup_{\ell \in [k]}\set{\bfa^\ell} \to B$ by
          \[
            f^k(a) = \begin{cases}
              f^{k-1}(a) & \text{if } a \in \bigcup_{\ell \in [k-1]}\set{\bfa^\ell} \\
              b^k_i & \text{if } a \in \set{\bfa^k} \setminus \bigcup_{\ell \in [k-1]}\set{\bfa^\ell} \text{ and }a = q^k_i.
            \end{cases}
          \]
        \EndIf
      \EndFor
      \State Reject.
    \EndProcedure
  \end{algorithmic}
\end{algorithm}

It is clear that \Call{HI-Test$_\bfA$}{} has a constant query complexity.
Also, note that if we have succeeded to define the map $f^k$, then it is a homomorphism from $\bfA$ to $\bfB$ by construction.
Hence, \Call{HI-Test$_\bfA$}{} always accepts when $\bfB$ is $\bfA$-free.\ifthenelse{\boolean{fullversion}}{
In the next section, we show that \Call{HI-Test$_\bfA$}{} rejects with a certain  probability if $\bfB$ is $\epsilon$-far from $\bfA$-freeness.
}{
In the full version, we show that \Call{HI-Test$_\bfA$}{} rejects with a certain  probability if $\bfB$ is $\epsilon$-far from $\bfA$-freeness.
}

\subsection{Soundness analysis}\label{subsec:analysis}

The goal of this subsection is to establish the following.
\begin{lemma}\label{lem:soundness}
  Suppose that $\bfB$ is $\epsilon$-far from $\bfA$-freeness for $\epsilon \in (0,1)$.
  Then, there exists $\delta = \delta^\bfA_{\ref{lem:soundness}}(\epsilon) > 0$ such that \textup{\Call{HI-Test$_\bfA$}{$B,\bfB$}} rejects with probability at least $\delta$.
  Here, we can choose $\delta^\bfA_{\ref{lem:soundness}}(\epsilon) = {\Omega\left(\epsilon/(m 2^{\bar{r}})\right)}^m$, where $\bar{r}=\max_{R \in \sigma} \ar(R)$.
\end{lemma}
In Section~\ref{subsubsec:reduction}, we introduce the notion of a tuple-disjoint $\bfA$-image, and show that we can restrict our attention to a relational structure constructed from tuple-disjoint $\bfA$-images.
Then, we analyze the rejection probability for it in Section~\ref{subsubsec:analysis-tuple-disjoint}.

\subsubsection{Restriction to the tuple-disjoint case}\label{subsubsec:reduction}

Let $\bfB$ be a multiset structure that is $\epsilon$-far from $\bfA$-freeness.
For a homomorphism $h:A \to B$ from $\bfA$ to $\bfB$, we define $h(\bfA)$ as the relational structure whose domain is $B$ and $R^{h(\bfA)} = \set{h(\bfa) \mid \bfa \in R^\bfA}$ for each $R \in \sigma$.
We call $h(\bfA)$ an \emph{$\bfA$-image} in $\bfB$.
We say that a multiset $\caI = \multiset{\bfI_1,\ldots,\bfI_k}$ of $\bfA$-images are \emph{tuple-disjoint} if, for each relation symbol $R \in \sigma$,
the multiset union of $R^{\bfI_1},\ldots,R^{\bfI_k}$ is a subset of $R^\bfB$.
For a set $\caI$ of tuple-disjoint $\bfA$-images in $\bfB$, we define $\bigcup \caI$ as the relational structure whose domain is $B$ and $R^{\bigcup \caI} = \bigcup_{\bfI \in \caI} R^{\bfI}$ for each $R \in \sigma$;
here, the union is taken as a multiset.

The following lemma states that we can analyze the rejection probability on $\bfB$ by analyzing the rejection probability on an instance induced by tuple-disjoint $\bfA$-images.
\begin{lemma}\label{lem:reduction}
  Suppose that $\bfB$ is a multiset structure,
 and $\caI$ is a set of tuple-disjoint $\bfA$-images in $\bfB$
 where $|\caI| \geq \alpha \| \bfB \|$ for a value $\alpha \in (0,1)$.
  Then, there is a constant $\kappa = \kappa^\bfA_{\ref{lem:reduction}}(\alpha) > 0$ and a subset $\caI' \subseteq \caI$ with $|\caI'| \geq |\caI|/2$ such that, if the probability that \textup{\Call{HI-Test$_\bfA$}{$B, \bigcup \caI'$}} rejects is $p$, then the probability that \textup{\Call{HI-Test$_\bfA$}{$B, \caB$}} rejects is at least $\kappa^{m} p$.
  Here, we can choose $\kappa^\bfA_{\ref{lem:reduction}}(\alpha) = \alpha / 2^{1+\max_{R \in \sigma} \ar(R)}$, where $\bar{r} = \max_{R \in \sigma} \ar(R)$.
\end{lemma}

\begin{proof}
  Consider the sum $\sum_{R, \bft}\deg_{R,\bfB}(\bft)$ over all pairs $(R,\bft)$ where $R \in \sigma$ and $\bft \in {(B \cup \set{*})}^{\ar(R)}$.
  For any relation symbol $R$, a tuple $\bfb \in R^\bfB$ is a completion of $2^{\ar(R)}$ tuples $\bft$.  Thus, we have
    $\sum_{R, \bft} \deg_{R,\bfB}(\bft) \leq C \|\bfB\|$
  for $C = 2^{\bar{r}}$.

  Set $\kappa = \alpha / 2C$ and initialize $\caI'$ to be $\caI$.
  Now, iterate the following process:
  if, for a pair $(R, \bft)$ of the described form, it holds that
  $\deg_{R,\bigcup \caI'}(\bft) < \kappa \cdot \deg_{R,\bfB}(\bft)$, remove each image $\bfI$ in $\caI'$ such that $R^\bfI$ includes a completion of $\bft$.
  As each pair $(R,\bft)$ is considered at most once in the process, at the end of the process, the total number of images removed is at most $\sum_{R,\bft} \kappa \cdot \deg_{R,\bfB}(\bft) \leq \kappa C\|\bfB\| = \alpha \|\bfB\| / 2$.
  In particular, $\caI'$ is non-empty.
  Moreover, we have $\deg_{R,\bigcup \caI'}(\bft) = 0$ or $\deg_{R,\bigcup \caI'}(\bft) \geq \kappa \cdot \deg_{R,\bfB}(\bft)$ for every
pair $(R, \bft)$.

  Now, we relate the probabilities that \Call{HI-Test$_\bfA$}{$B, \bfB$} and \Call{HI-Test$_\bfA$}{$B, \bigcup \caI'$} reject.
  To this end, fix a sequence of tuples $\bfb^1 \in R_1^{\bigcup \caI'},\ldots,\bfb^m \in R_m^{\bigcup \caI'}$ that forms a homomorphism $g:\bfA \to \bigcup \caI'$, that is, $g(\bfa^i) = \bfb^i$ for each $i \in [m]$.
  For $k \in [m]$, let $p_{\bfB}(\bfb^1,\ldots,\bfb^k)$ and $p_{\bigcup \caI'}(\bfb^1,\ldots,\bfb^k)$ be probabilities that \Call{HI-Test$_\bfA$}{$B, \bfB$} and \Call{HI-Test$_\bfA$}{$B, \bigcup \caI'$} obtain the sequence $\bfb^1,\ldots,\bfb^k$, respectively.
  Now, we show that $p_{\bfB}(\bfb^1,\ldots,\bfb^k)$ is at least $\kappa^k$ times $p_{\bigcup \caI'}(\bfb^1,\ldots,\bfb^k)$.
  Considering the case $k = m$ and summing up the inequalities over the sequence $\bfb^1,\ldots,\bfb^m$, we get the claim.

  Suppose that we have obtained $\bfb^1,\ldots,\bfb^{k-1}$ through the oracle (both in \Call{HI-Test$_\bfA$}{$B, \bfB$} and \Call{HI-Test$_\bfA$}{$B, \bigcup \caI'$}).
  Then, the probability that \Call{HI-Test$_\bfA$}{$B, \bfB$} obtains $\bfb^k$ in the $k$-th stage is
  \[
    \frac{1}{\deg_{R_k,\bfB}(\bfq^k)}
    =
    \frac{\kappa}{\kappa \deg_{R_k,\bfB}(\bfq^k)}
    \geq
    \frac{\kappa}{\deg_{R_k,\bigcup \caI'}(\bfq^k)},
  \]
  which is $\kappa$ times the probability that \Call{HI-Test$_\bfA$}{$B, \bigcup \caI'$} obtains $\bfb^k$.
  Note that $\deg_{R_k,\bigcup \caI'}(\bfq^k)$ is not zero because of the existence of $\bfb^k \in R_k^{\bigcup \caI'}$.
  By induction on $k$, it follows that
   $p_{\bfB}(\bfb^1,\ldots,\bfb^k) \geq \kappa^k \cdot  p_{\bigcup \caI'}(\bfb^1,\ldots,\bfb^k)$.
\end{proof}

In light of Lemma~\ref{lem:reduction}, it suffices to
find a set of tuple-disjoint $\bfA$-images,
and then analyze the rejection probability thereof.

\subsubsection{Analysis on a relational structure constructed from tuple-disjoint $\bfA$-images}\label{subsubsec:analysis-tuple-disjoint}

Now, we construct a large multiset of tuple-disjoint $\bfA$-images in $\bfB$.
Let $\bfB$ be a multiset structure that is $\epsilon$-far from $\bfA$-free,
and let $G$ be a maximal multiset of homomorphisms from $\bfA$ to $\bfB$ such that the multiset of the $\bfA$-images
$\multiset{ g(\bfA) \mid g \in G }$ are tuple-disjoint.

Let $G' \subseteq G$ be the subset with $|G'| \geq |G|/2$ given by applying Lemma~\ref{lem:reduction} on $G$.
Let $\bfB' = (B, \set{R^{\bfB'}}_{R\in \sigma})$ be the
multiset structure obtained by taking the union of the images $\multiset{g(\bfA) \mid g \in G'}$,
that is, the structure where $R^{\bfB'} = \cup_{g \in G'} R^{g(\bfA)}$
for each $R \in \sigma$.
For $k \in \{ 0, \ldots, m \}$, we say that the algorithm is \emph{truly $k$-successful} if, for each $\ell \in [k]$,
there exists $g \in G'$ such that $\bfb^\ell = g(\bfa^\ell)$,
where $\bfb^\ell$ is the answer
to the $\ell$-th query.
Let $T_k$ denote the event of being truly $k$-successful.

\begin{lemma}\label{lem:truly-k-successful}
  Suppose we run \textup{\Call{HI-Test$_\bfA$}{$B,\bfB'$}}.
  Then, it holds that
  $\Pr\left[T_k \mid T_{k-1}\right] \geq \frac{1}{m}$ %
  for every $k \in [m]$.
\end{lemma}

\begin{lemma}\label{lem:J-large}
  We have $|G| >\epsilon\|\bfB\|/m$.
\end{lemma}
\begin{proof}
  We could remove all $\bfA$-images in $\bfB$ by removing all tuples in $\multiset{g(\bfa) \mid g \in G, R \in \sigma, \bfa \in R^\bfA}$.
  This means that we can delete all homomorphisms from $\bfA$ to $\bfB$ by removing at most $m \cdot |G|$ tuples from $\bfB$.
  From the $\epsilon$-farness of $\bfB$, we have $m \cdot |G| > \epsilon \|\bfB\|$, which yields the claim.
\end{proof}

\begin{lemma}\label{lem:f-g-equal}
  After the $k$-th stage of \textup{\Call{HI-Test$_\bfA$}{$B,\bfB'$}},
  for any map $h: \set{\bfa^k}\to B$, we have
  \[
    \Pr\Bigl[f^k|_{\set{\bfa^k}} = h \mid T_k\Bigr] =
    \frac{\#\multiset{g \in G' \mid g\text{ extends }h}}{|G'|}.
  \]
\end{lemma}
\begin{proof}
  We prove this by induction.
  It is clear for $k = 1$, since each of the tuples in $\multiset{g(\bfa^1) \mid g \in G'}$ is chosen with equal probability.

  For $k > 1$, let $\ell < k$ denote an index
  such that $P_k := \set{\bfa^k} \cap \bigcup_{\ell' \in [k-1]} \set{\bfa^{\ell'}} \subseteq \{ \bfa^\ell \}$;
  such an index $\ell$ exists due to $\bfA$ being $\alpha$-acyclic.
  From the induction hypothesis on the $\ell$-th stage, we have (by summing)
  that for any map $h: P_k \to B$,
  \begin{align*}
    & \Pr\Bigl[f^\ell|_{P_k} = h \mid T_k \Bigr]
    = \Pr\Bigl[f^\ell|_{P_k} = h \mid T_\ell \Bigr] \\
    & = \frac{\#\multiset{g \in G'\mid g\text{ extends }h}}{|G'|}.
  \end{align*}

We then have
  \begin{align*}
    & \Pr\Bigl[f^k|_{\set{\bfa^k}} = h \mid T_k \Bigr] \\
    & = \Pr\Bigl[f^\ell|_{P_k} = h|_{P_k} \mid T_k\Bigr] \\
    & \quad \times \Pr\Bigl[f^k|_{\set{\bfa^k}\setminus P_k} = h|_{\set{\bfa^k} \setminus P_k} \mid T_k \wedge f^\ell|_{P_k} = h|_{P_k}  \Bigr] \\
    & = \Pr\Bigl[f^\ell|_{P_k} = h|_{P_k} \mid T_\ell \Bigr] \\
    & \quad \times \Pr\Bigl[f^k|_{\set{\bfa^k}\setminus P_k} = h|_{\set{\bfa^k} \setminus P_k} \mid T_k \wedge f^\ell|_{P_k} = h|_{P_k}  \Bigr]. \\
  \end{align*}
  The latter quantity is equal to the product of 
    $$\#\multiset{g \in G' \mid g\text{ extends }h|_{P_k}} / |G'|$$ 
and
    $$\#\multiset{g \in G' \mid g \text{ extends }h} / \#\multiset{g \in G' \mid g \text{ extends }h|_{P_k}}$$
which yields the lemma.
\end{proof}

\begin{proof}[Proof of Lemma~\ref{lem:truly-k-successful}]
  Let $P_k := \set{\bfa^k} \cap \bigcup_{\ell' \in [k-1]} \set{\bfa^{\ell'}}$ be the set of elements previously appeared.

  Suppose that $P_k = \set{\bfa^k}$.
  Then, there exists $\ell \in [k-1]$ such that $\set{\bfa^k} \subseteq \set{\bfa^\ell}$.
  Then, as $\bfq^k = f^\ell(\bfa^\ell)|_{\set{\bfa^k}} = g(\bfa^\ell)|_{\set{\bfa^k}}$ for some homomorphism $g \in G'$, we have $\bfq^k = g(\bfa^k)$ and hence the process is truly $k$-successful.

  Suppose that $P_k \subsetneq \set{\bfa^k}$.
  We define $Q \subseteq {(B \cup \set{*})}^{\ar(R_k)}$ as the set of tuples $\bfq$ such that (i) $\bfq(a) \neq *$ for $a \in P_k$, (ii) $\bfq(a) = *$ for $a \in \set{\bfa^k} \setminus P_k$, and (iii) $\bfq$ has a completion in $\multiset{g(\bfa^k) \mid g \in G'}$.
  For $\bfq \in Q$, we define $n_\bfq$ as the number of completions $\bfb$ of $\bfq$ such that $\bfb = g(\bfa^k)$ for some $g \in G'$.
  Note that $\sum_{\bfq \in Q}n_\bfq= |G'|$.
  For $\bfq \in Q$, we define $m_\bfq$ as the total number of completions $\bfb \in R_k^{\bfB'}$ of $\bfq$, that is, $\bfb = g(\bfa^\ell)$ for some $\ell \in [m]$ with $R_\ell = R_k$.
  Note that $\sum_{\bfq \in Q} m_\bfq \leq m \cdot \sum_{\bfq \in Q}n_{\bfq} = m|G'|$ because $\bfA$-images of homomorphisms in $G'$ are tuple-disjoint.
  Also, $m_\bfq \geq n_\bfq > 0$ for any $\bfq \in Q$ from the definition of $Q$.

  Let $\ell \in [k-1]$ be such that $P_k \subseteq \set{\bfa^\ell}$.
  For $\bfq\in Q$, define $h_\bfq:P_k \to B$ as $h(a) = \bfq(a)$ for every $a\in P_k$.
  By Lemma~\ref{lem:f-g-equal}, we have
  \begin{align*}
    & \Pr\Bigl[\bfq^k = \bfq \mid T_{k-1} \Bigr]
    =
    \Pr\Bigl[f^{k-1}|_{P_k} = h_\bfq \mid T_{k-1} \Bigr] \\
    & =
    \frac{\#\multiset{g \in G' \mid g\text{ extends }h_{\bfq}} }{|G'|}
    =
    \frac{n_\bfq}{|G'|}.
  \end{align*}
  Hence, we have
  \begin{align}
    \Pr\Bigl[T_k \mid T_{k-1} \Bigr]
    =
    \sum_{\bfq \in Q} \frac{n_\bfq}{|G'|}\frac{n_\bfq}{m_\bfq}. \label{eq:target}
  \end{align}
  Now we look at~\eqref{eq:target} as a function of $\bmm = \set{m_\bfq}_{\bfq\in Q}$ and write it as $p(\bmm)$.
  Then, to get a lower bound on~\eqref{eq:target}, we extend the domain of $p$ from $\bbN^Q$ to $\bbR_+^Q$ and minimize it subject to $\sum_{\bfq \in Q}m_\bfq \leq m \cdot \sum_{\bfq \in Q}n_{\bfq} $.
  Notice that
  \[
    {(\nabla p)}_\bfq = \begin{cases}
    -\frac{n_\bfq^2}{|G'|m_\bfq^2} & \text{if } m_\bfq > 0,\\
    0 & \text{otherwise}.
    \end{cases}
  \]
  This in particular means that when $m_{\bfq} = 0$ for some $\bfq \in Q$, we can decrease the value of $p$ by increasing $m_{\bfq}$ while decreasing another element.
  Hence, an optimizer $\bmm^* = \set{m^*_\bfq}_{\bfq\in Q}$ to~\eqref{eq:target} is not on the boundary of the domain $\bbR_+^Q$.

  Then, we have for any $\bfq, \bfq' \in Q$ that
  \[
    \nabla p(\bmm^*) \cdot (\bme_{\bfq} - \bme_{\bfq'}) = \frac{1}{|G'|}\Bigl(\frac{n_{\bfq'}^2}{{(m^*_{\bfq'})}^2} -\frac{n_{\bfq}^2}{{(m^*_{\bfq})}^2}\Bigr),
  \]
  where $\bme_\bfq$ is the characteristic vector with one in the $\bfq$-th element and zeros elsewhere.
  As $\bmm^*$ satisfies $\nabla f(\bmm^*) \cdot (\bme_{\bfq} - \bme_{\bfq'}) = 0$ for all $\bfq, \bfq' \in B$, $n_\bfq/m_\bfq$ is constant regardless of the choice of $\bfq$.
  It follows that
  \[
    p(\bmm^*) \geq \sum_{\bfq\in Q} \frac{n_\bfq}{|G'|} \frac{n_\bfq}{m n_\bfq} = \frac{1}{m},
  \]
  which implies the claim.
\end{proof}

\begin{corollary}\label{cor:rejection-probability-on-G}
  The rejection probability of \textup{\Call{HI-Test$_\bfA$}{$B,\bfB'$}} is at least $1/m^{m}$.
\end{corollary}
\begin{proof}
  By Lemma~\ref{lem:truly-k-successful}, the probability that the algorithm is truly $m$-successful and hence finds a homomorphism from $\bfA$ to $\bfB'$ is at least $1/m^m$.
\end{proof}

The proof of Lemma~\ref{lem:soundness} is immediate from Lemma~\ref{lem:reduction} and Corollary~\ref{cor:rejection-probability-on-G}.

With Lemma~\ref{lem:soundness}, we can easily get a tester for $\bfA$-freeness:
\begin{theorem}\label{the:alpha-acyclic}
  There exists a one-sided error tester for $\bfA$-freeness with a constant number of queries.
  The query complexity is ${O(m 2^{\bar{r}}/\epsilon)}^m$, where $\bar{r}=\max_{R \in \sigma} \ar(R)$.
\end{theorem}
\begin{proof}
  Our algorithm runs \textup{\Call{HI-Test$_\bfA$}{$B,\bfB$}} $\Theta(1/\delta)$ many times for $\delta = \delta^\bfA_{\ref{lem:soundness}}(\epsilon)$, and accepts if and only if all the runs accept.

  If $\bfB$ is $\bfA$-free, then \Call{HI-Test$_\bfA$}{} always accepts,
  and the algorithm also always accepts.
  If $\bfB$ is $\epsilon$-far from $\bfA$-freeness, then the rejection probability of the algorithm is at least $1 - {(1 - \delta)}^{\Theta(1/\delta)} \geq \frac{2}{3}$ by setting the hidden constant to be large enough.
\end{proof}
Theorem~\ref{the:alpha-acyclic} gives the ``if'' part of Theorem~\ref{the:main} because, if the core $\bfA'$ of $\bfA$ is $\alpha$-acyclic, then we can test $\bfA$-freeness just by testing $\bfA'$-freeness.

Suppose $\bfA$ represents an undirected graph $H$ as in Section~\ref{subsect:model-relationship}.
Then, as $\bar{r}=2$, the query complexity is ${O(m/\epsilon)}^m$, where $m$ is the number of edges in $H$.
This bound matches the previous result for testing $k$-path-freeness~\cite{Iwama:2018bs}.

\subsection{Injective homomorphism inadmissibility}\label{subsec:injective}

In this section, we discuss how to test injective homomorphism inadmissibility.
The modification to Algorithm~\ref{alg:alpha-cyclic} is simple: we check whether the obtained homomorphism is injective at the end of the algorithm.
See Algorithm~\ref{alg:alpha-cyclic-injective}, called \Call{IHI-Test$_\bfA$}{} for details.
It is clear that \Call{IHI-Test$_\bfA$}{} always accepts $\bfB$ that is injective $\bfA$-free and that the query complexity is constant.
Hence, we analyze its rejection probability when $\bfB$ is $\epsilon$-far from injective $\bfA$-freeness.

\begin{algorithm}[t!]
  \caption{}\label{alg:alpha-cyclic-injective}
  \begin{algorithmic}[1]
    \Procedure{IHI-Test$_{\bfA}$}{$B$, $\bfB$}
      \Comment{$B$ is a domain, $\bfB$ is a multiset structure with the domain $B$ and the same vocabulary as $\bfA$ provided through a query access.}
      \If{\Call{HI-Test$_\bfA$}{$B$, $\bfB$} accepts (by claiming that $\bfB$ is $\bfA$-free)}
        \State Accept.
      \Else
        \State Let $g:A \to B$ be the obtained homomorphism.
        \If{$g$ is not injective}
          \State Accept.
        \Else
          \State Reject.
        \EndIf
      \EndIf
    \EndProcedure
  \end{algorithmic}
\end{algorithm}

The analysis of \Call{IHI-Test$_\bfA$}{} is very similar to the one of \Call{HI-Test$_\bfA$}{}.
We use the reduction given by Lemma~\ref{lem:reduction} and focus on a relational structure constructed from tuple-disjoint $\bfA$-images.
However, we carefully choose the images so that we get an injective homomorphism with high probability.

Now, we construct a large set of tuple-disjoint $\bfA$-images.
Let $\bfB$ be a multiset structure $\epsilon$-far from injective $\bfA$-freeness, and let $G$ be a maximal set of injective homomorphisms from $\bfA$ to $\bfB$ such that the multiset of $\bfA$-images $\multiset{g(\bfA) \mid g \in G}$ is tuple-disjoint.
\begin{lemma}\label{lem:J-large-injective}
  We have $|G| >\epsilon\|\bfB\|/m$.
\end{lemma}
\begin{proof}
  We could remove all $\bfA$-images in $\bfB$ by removing all tuples in $\multiset{g(\bfa) \mid g \in G, R \in \sigma, \bfa \in R^\bfA}$.
  This means that we can delete all injective homomorphisms from $\bfA$ to $\bfB$ by removing at most $m  |G|$ tuples from $\bfB$.
  From the $\epsilon$-farness of $\bfB$, we have $m |G| > \epsilon \|\bfB\|$, which yields the claim.
\end{proof}

Let $\phi:B \to A$ be a map, which we call a \emph{coloring} of $B$ with $|A|$ colors.
Then, we say that a homomorphism $g:A \to B$ \emph{admits the coloring $\phi$} if $\phi(g(a))=a$.
We define $G_\phi$ as the multiset of injective homomorphisms in $G$ that admit $\phi$.
\begin{lemma}\label{lem:good-coloring}
  There exists a coloring $\phi:B \to A$ such that $|G_\phi| \geq |G|/|A|^{|A|}$.
\end{lemma}
\begin{proof}
  Suppose that a coloring $\psi:B \to A$ is randomly chosen, that is, $\Pr[\psi(b)] = 1/|A|$ for every $b \in B$ independently from others.
  Then, the probability that an injective homomorphism $g:A \to B$ admits the coloring $\psi$ is $1/|A|^{|A|}$.
  It follows that the expected number of homomorphisms in $G$ that admit $\psi$ is $|G|/|A|^{|A|}$.
  Hence, there exists a coloring $\phi:B \to A$ with $|G_\phi| \geq |G|/|A|^{|A|}$.
\end{proof}

Fix a coloring $\phi: B \to A$ with $|G_\phi| \geq |G| / |A|^{|A|}$, given by Lemma~\ref{lem:good-coloring}.
Let $G'_\phi \subseteq G$ be the subset with $|G'_\phi| \geq |G_\phi|/2$ given by applying Lemma~\ref{lem:reduction} on $G_\phi$.
Then, we define a multiset structure $\bfB'_\phi$ and the notion of being truly $k$-successful using $G'_\phi$, as in Section~\ref{subsubsec:analysis-tuple-disjoint}.

The advantage of working with $G'_\phi$ is that, if the algorithm is truly $m$-successful with respect to $G'_\phi$, then the obtained homomorphism is injective because any homomorphism $g$ obtained when truly $m$-successful satisfies $\phi(g(a)) = a$ for every $a \in A$ and hence $g(a)$'s are distinct.

Modifying the proof of Corollary~\ref{cor:rejection-probability-on-G} by replacing $\bfB'$ with $\bfB'_\phi$, we obtain the following:
\begin{corollary}\label{cor:rejection-probability-on-G-injective}
  \textup{\Call{HI-Test$_\bfA$}{$B,\bfB'_\phi$}} finds an injective homomorphism with probability at least $1/m^{m}$.
\end{corollary}

\begin{lemma}\label{lem:soundness-injective}
  Suppose that the multiset structure $\bfB$ is $\epsilon$-far from injective $\bfA$-freeness for $\epsilon \in (0,1)$.
  Then, there exists $\delta = \delta^\bfA_{\ref{lem:soundness-injective}}(\epsilon) > 0$ such that \textup{\Call{IHI-Test$_\bfA$}{$B,\bfB$}} rejects with probability at least $\delta$.
\end{lemma}
\begin{proof}
  It is immediate from Lemma~\ref{lem:reduction} and Corollary~\ref{cor:rejection-probability-on-G-injective} that there exists $\delta = \delta(\epsilon) > 0$ such that \textup{\Call{HI-Test$_\bfA$}{$B,\bfB$}} rejects and returns an injective homomorphism $g:A \to B$ from $A$ to $\bfB'_\phi$ with probability at least $\delta$.
  As $g$ is injective, \textup{\Call{IHI-Test$_\bfA$}{$B,\bfB$}} also rejects with probability at least $\delta$.
\end{proof}

\begin{theorem}\label{the:alpha-acyclic-injective}
When $\bfA$ is $\alpha$-acyclic,
  there exists a one-sided error tester for injective $\bfA$-freeness with a constant number of queries.
\end{theorem}
\begin{proof}
  Our algorithm run \textup{\Call{IHI-Test$_\bfA$}{$B,\bfB$}} $\Theta(1/\delta)$ many times for $\delta = \delta^\bfA_{\ref{lem:soundness-injective}}(\epsilon)$, and accepts if and only if all the runs accept.

  If there is no injective homomorphism from $\bfA$ to $\bfB$, then \Call{IHI-Test$_\bfA$}{} always accepts, and the algorithm also always accept.
  If $\bfB$ is $\epsilon$-far from injective $\bfA$-freeness, then the rejection probability of the algorithm is at least $1 - {(1 - \delta)}^{\Theta(1/\delta)} \geq \frac{2}{3}$ by setting the hidden constant to be large enough.
\end{proof}

\section{Lower Bound}\label{sec:lower-bound}

In this section, we establish
the following lower bound result.

\begin{theorem}\label{thm:lower-bound}
Suppose that $\bfA$ is a relational structure whose core
is not $\alpha$-acyclic; then,
testing $\bfA$-freeness
is not constant-query testable with one-sided error.
\end{theorem}

In order to establish this lower bound, we will make use of the following notion of reduction.

\begin{definition}\label{def:reduction}
Let $\bfA_1$, $\bfA_2$ be structures.
We say that
testing $\bfA_1$-freeness
\emph{reduces} to
testing $\bfA_2$-freeness
if there exists a map $f$ sending
each multiset structure $\bfB_1$
to a multiset structure $\bfB_2 = f(\bfB_1)$
such that the following hold:
\begin{itemize}
\itemsep=0pt
\item[(i)] For each structure $\bfB_1$, if $\bfB_1$ is $\bfA_1$-free,
then $\bfB_2$ is $\bfA_2$-free.

\item[(ii)] For each $\epsilon_1 \in (0,1)$,
there exists $\epsilon_2 \in (0,1)$
such that, for each structure $\bfB_1$,
 if
$\bfB_1$ is $\epsilon_1$-far from being $\bfA_1$-free,
then
$\bfB_2$ is $\epsilon_2$-far from being $\bfA_2$-free.

\item[(iii)] There exists a constant $C \geq 1$
such that, for each structure $\bfB_1$,
the answer to a query to the completion oracle
for $\bfB_2$ can be computed from the answers to $C$ many queries
to the completion oracle for $\bfB_1$.

\item[(iv)] There exists a function $g(n) \in O(n)$ such that,
for each structure $\bfB_1$, it holds that
$|B_2| \leq g(|B_1|)$.
\end{itemize}

\end{definition}

We will make tacit use of the following straightforwardly verified facts.

\begin{proposition}
Suppose that $\bfA_1$, $\bfA_2$ are structures such that
testing $\bfA_1$-freeness
reduces to
testing $\bfA_2$-freeness.
\begin{itemize}
\item If $\bfA_2$-freeness is constant-query testable with one-sided error,
then so is $\bfA_1$-freeness.

\item If $k \geq 1$ is a constant such that
 $\bfA_2$-freeness is testable in $O(n^{1/k})$ queries with one-sided error,
 then $\bfA_1$-freeness is also so testable.
\end{itemize}
\end{proposition}

Towards proving Theorem~\ref{thm:lower-bound},
we next present two reductions.  The first
allows us to restrict attention to symbol-distinct structures,
and the second allows us to pass to induced hypergraphs.
Here, we say that a structure is \emph{symbol-distinct} if each relation has exactly one tuple.

The full proofs of the following two theorems are included in the appendix.

\begin{theorem}\label{thm:reduction-symbol-distinct}
For each structure $\bfA$ that is a core,
there exists a symbol-distinct structure $\bfA'$
such that
testing $\bfA'$-freeness
reduces to
testing $\bfA$-freeness,
and such that $\bfA$ and $\bfA'$ share the same hypergraph, that is, $G_\bfA $ and $G_{\bfA'}$ are isomorphic.
\end{theorem}

The idea behind the proof of Theorem~\ref{thm:reduction-symbol-distinct}
is as follows.  The structure $\bfA'$ is defined
on a vocabulary $\sigma'$ having a relation symbol $R_{\bft}$ interpreted as $\{ \bft \}$,
for every relation symbol $R$ of the vocabulary $\sigma$ of $\bfA$ and for each $\bft \in R^{\bfA}$.
To reduce from testing $\bfA'$-freeness to testing $\bfA$-freeness,
a structure $\bfB'$ with vocabulary $\sigma'$
is translated to a structure $\bfB$ with vocabulary $\sigma$  by
marking the elements of $R_{\bft}^{\bfB'}$ with $\bft$ and then placing them in $R^{\bfB}$.
Establishing the correctness of the reduction uses the assumption that $\bfA$ is a core.

\begin{theorem}\label{thm:reduction-induced-hypergraph}
Let $\bfA$ be a symbol-distinct structure
whose hypergraph is $G_{\bfA}$,
and let $S \subseteq A$.
There exists a symbol-distinct structure $\bfA'$
whose hypergraph is $G_{\bfA}[S]$
such that testing $\bfA'$-freeness
reduces to
testing $\bfA$-freeness.
\end{theorem}

On a high level, Theorem~\ref{thm:reduction-induced-hypergraph}
is established in the following way.
For each relation symbol $R$ in the signature $\sigma$ of $\bfA$,
the structure $\bfA'$ has a relation whenever the tuple $(a_1, \ldots, a_r) \in R^{\bfA}$
contains an entry from $S$; that relation is interpreted as the relation having one tuple
whose entries are the values among $a_1, \ldots, a_r$ that are contained in $S$;
the other coordinates are effectively removed.
To reduce from testing $\bfA'$-freeness to testing $\bfA$-freeness,
a structure $\bfB'$ over signature $\sigma'$ is translated to a structure $\bfB$
over signature $\sigma$ essentially by filling out the coordinates
that were removed, in each relation, with a value $+$ that is fresh in that it is not contained in $B$.
In this way, any mapping $f: S \to B'$ can be naturally extended
to a mapping $f': A \to B$ where each element of $A \setminus S$ is sent to $+$;
this correspondence has the property that $f$ is a homomorphism
from $\bfA'$ to $\bfB'$ if and only if $f'$ is a homomorphism from $\bfA$ to $\bfB$.

We now give a proof of
Theorem~\ref{thm:lower-bound} based on the following two theorems, whose proofs are deferred to Sections~\ref{subsec:cycles} and~\ref{subsec:simplices}.

\begin{theorem}\label{the:cycle-freeness}
  Let $\bfA$ be a symbol-distinct relational structure such that $\caM(G_\bfA)$ is a cycle.
  Then, testing $\bfA$-freeness requires $\Omega(\sqrt{n})$ queries, where $n$ is the domain size of the input structure.
\end{theorem}

\begin{theorem}\label{the:simplex-freeness}
  Let $\bfA$ be a symbol-distinct relational structure such that $\caM(G_\bfA)$ is a $k$-simplex for $k \geq 2$.
  Then, testing $\bfA$-freeness with one-sided error requires $\Omega(n^{1/k})$ queries, where $n$ is the domain size of the input structure.
\end{theorem}

\begin{proof}[Proof of Theorem~\ref{thm:lower-bound}]
Let $\bfC$ be the core of $\bfA$.
Since $\bfA$ and $\bfC$ are homomorphically equivalent
(that is, there exist homomorphisms from each to the other),
the properties of $\bfA$-freeness and $\bfC$-freeness are equivalent.
Hence, it suffices to prove the lower bound
for testing $\bfC$-freeness.
By Theorem~\ref{thm:reduction-symbol-distinct},
there exists a symbol-distinct structure $\bfC'$
whose hypergraph $G$ is not $\alpha$-acyclic,
and such that it suffices to prove the lower bound
for testing $\bfC'$-freeness.
We now invoke Theorem~\ref{the:characterization-of-alpha-acyclic};
by this theorem, there exists a subset $S$ of the domain of $\bfC'$,
with $|S| \geq 3$,
such that $\caM(G[S])$ is either a cycle or a simplex.
By Theorem~\ref{thm:reduction-induced-hypergraph},
there exists a symbol-distinct structure $\bfC''$
whose hypergraph is $H = G[S]$ such that
it suffices to prove the lower bound
for testing $\bfC''$-freeness.
We have that $\caM(H)$ is either a cycle or a simplex.
The result follows from
Theorems~\ref{the:cycle-freeness}
and~\ref{the:simplex-freeness}.
\end{proof}

\subsection{Lower Bounds for Cycles}\label{subsec:cycles}

In this section, we prove Theorem~\ref{the:cycle-freeness}.
We use a reduction from testing triangle-freeness, which is hard as in the following theorem:
\begin{theorem}[\cite{Alon:2008gn}]\label{the:triangle-freeness}
  Testing triangle-freeness in the general graph model requires $\Omega(\sqrt{n})$ queries, where $n$ is the number of vertices in the input graph.
\end{theorem}
The model studied in~\cite{Alon:2008gn} did not consider queries of the form $(*,*)$.
However, as the hardness is shown on a distribution over $d$-regular graphs for a fixed $d$, we can answer to the size query $(*,*)$ by returning $dn/2$ and to the completion query $(*,*)$ by sampling a vertex uniformly at random and returning a random edge incident to it using the random neighbor oracle.

\begin{proof}[Proof of Theorem~\ref{the:cycle-freeness}]
  Let $A = \set{a_1,\ldots,a_k}$, where $k \geq 3$.
  From the assumption, every relation in $\bfA$ is unary or binary, and any binary relation of $\bfA$ has a single tuple of the form $(a_i,a_{i+1})$ for $i \in [m]$, where $a_{k+1}=a_1$.

  Now, we reduce testing triangle-freeness in the general graph model to testing $\bfA$-freeness.
  Given a query access to a (regular) graph $G=(V,E)$, we construct a multiset structure $\bfB$ on the domain $V$ as follows.
  For each unary relation $R^\bfA$, we define $R^\bfB$ as the set containing $2|E|/|V|$ many copies of $(v)$ for each $v \in V$.
  Note that, as $G$ is regular, $|V|$ always divides $2|E|$.
  For each binary relation of the form $R^\bfA = \set{(a_i,a_{i+1})}$ for $i \in [3]$, we define $R^\bfB = \set{(u,v),(v,u) \mid \set{u,v} \in E}$.
  For each binary relation of the form $R^\bfA = \set{(a_i,a_{i+1})}$ for $i \in \set{4,\ldots,k}$, we define $R^\bfB$ as the set containing $2|E|/|V|$ many copies of $(v,v)$ for each $v \in V$.
  Now, note that every relation contains exactly $2|E|$ tuples.
  Hence, $\|\bfB\| = 2\|\bfA\| \cdot |E|$.

  Now, we check the reduction above satisfies the conditions of Definition~\ref{def:reduction}.
  If $G$ is triangle-free, then clearly $\bfB$ is $\bfA$-free, which confirms (i).
  Suppose that $G$ is $\epsilon$-far from $\bfA$-freeness whereas $\bfB$ is $\delta$-close to $\bfA$-freeness.
  Removing $\delta\|\bfB\|$ tuples from relations in $\bfB$ corresponds to removing at most $2\delta \|\bfA\|\cdot |E| = 2\delta \|\bfA\| \cdot |E|$ edges and at most $2\delta \|\bfA\| \cdot |E| \cdot |V|/2|E| = \delta \|\bfA\| \cdot |V|$ vertices from $G$.
  Recalling that $G$ is regular, it results in removing at most $4\delta \|\bfA\| \cdot |E|$ edges.
  Hence, we must have $4\delta \|\bfA\| \geq \epsilon$, which confirms (ii).
  It is easy to see that we can answer to a query to the oracle for $\bfB$ using a constant number of queries to the oracle for $G$, which confirms (iii).
  Hence, testing $\bfA$-freeness requires $\Omega(\sqrt{n})$ queries by Theorem~\ref{the:triangle-freeness}.
\end{proof}

\subsection{Lower Bounds for Simplices}\label{subsec:simplices}

In this section, we prove Theorem~\ref{the:simplex-freeness}.
The main ingredient of our proof is a hardness for testing simplex-freeness of hypergraphs.
To state the hardness, we first introduce the general hypergraph model, which is an extension of the general graph model.
In this model, an algorithm is given an integer $n$ and a query access to a $k$-uniform hypergraph $G=(V,\caE)$ on $n$ vertices.
There are two types of queries, called \emph{size queries} and \emph{completion queries}.
Each type of query receives a subset $Q \subseteq V$ of size at most $k$.
A hyperedge $S \in \caE$ is called a \emph{completion} of $Q$ in $G$ if $S \cap Q = Q$.
Then, the answer to the size query $Q$ is the number of completions of $Q$ in $G$, and the answer to the completion query $Q$ is a random completion of $Q$ in $G$ if exists and a special symbol $\bot$ otherwise.
In particular, if $Q$ is empty, then it returns a random hyperedge (unless $G$ is empty), and if $Q$ is of size $k$, then it returns whether $G$ has the hyperedge $Q$.
A hypergraph $G=(V,\caE)$ is called \emph{$\epsilon$-far from simplex-freeness} if we must remove more than $\epsilon |E|$ hyperedges so that the resulting hypergraph has no simplex as its subgraph.
A \emph{tester} for simplex-freeness is an algorithm that, given an integer $n$, a query access to a graph $G=(V,\caE)$, and an error parameter $\epsilon \in (0,1)$, accepts with probability at least $2/3$ if $G$ is simplex-free and rejects with probability at least $2/3$ if $G$ is $\epsilon$-far from simplex-freeness.
Then, we show the following lower bound for testing simplex-freeness.
\begin{theorem}\label{the:hypergraph-simplex-freeness}
  For any $k \geq 2$, testing $k$-simplex in a $(k+1)$-uniform hypergraph with one-sided error requires $\Omega(n^{1/k})$ queries, where $n$ is the number of vertices in the input hypergraph.
\end{theorem}

In Section~\ref{subsubsec:simplices-distribution}, we provide a distribution of $k$-uniform hypergraphs that have many edge-disjoint $k$-simplices.
In Section~\ref{subsubsec:simplices-one-sided-error}, we show that any one-sided error tester on a hypergraph sampled from the distribution requires $\Omega(n^{1/k})$ queries.
We note that the proof of Lemma~\ref{the:hypergraph-simplex-freeness} follows the line of the proof of Theorem~\ref{the:triangle-freeness}, but is more cumbersome.
The proof of Theorem~\ref{the:simplex-freeness} is provided in Section~\ref{subsec:simplex-freeness}.
Missing proofs are given in Appendix~\ref{sec:simplices-appendix}.

\subsubsection{Lower-bound distribution}\label{subsubsec:simplices-distribution}

Let $\caD_{n,k}$ be a distribution over $k$-uniform hypergraphs with $n$ vertices and degree $d = \frac{k}{k+1}\cdot{(\frac{n}{k+1})}^{k-1-1/k}$ defined as follows.
A hypergraph is generated by first partitioning the vertices into equal-size subsets of size $\bar{n} = n/(k+1)$, denoted $V_1,\ldots,V_{k+1}$.
Next, for each subset $\set{i_1,\ldots,i_k} \subseteq [k+1]$ of size $k$, we select $\bar{d} = d/k = \bar{n}^{k-1-1/k}/(k+1)$ random perfect $k$-dimensional matchings among $V_{i_1},\ldots,V_{i_k}$.
In what follows, we assume that $\bar{n}$ is sufficiently large.

\begin{lemma}\label{lem:hypergraph-far}
  For any $n$ and $k$, a hypergraph sampled uniformly from $\caD_{n,k}$ is $\Omega(1)$-far from simplex-freeness with probability at least $\delta_{\ref{lem:hypergraph-far}}(k) = \frac{1}{4{(k+1)}^k}$.
\end{lemma}

\subsubsection{Lower bound for one-sided error testers}\label{subsubsec:simplices-one-sided-error}

The goal of this section is proving Theorem~\ref{the:hypergraph-simplex-freeness}.
Let $\caA$ be any one-sided error tester for simplex-freeness.
We can assume the probability that $\caA$ errs is an arbitrarily small constant, say, $\delta_{\ref{lem:hypergraph-far}}(k)/2$ (Otherwise, we can run $\caA$ many times to reduce the failure probability).
If $\caA$ views a simplex in the tested hypergraph, then clearly it can reject the hypergraph.
However, since $\caA$ is a one-sided error tester, if it terminates before viewing a simplex, then it must accept.
Suppose that we run $\caA$ on a hypergraph chosen according to $\caD_{n,k}$ introduced in Section~\ref{subsubsec:simplices-distribution}, with some (sufficiently small) constant $\epsilon$.
Since we have shown in Lemma~\ref{lem:hypergraph-far} that, with  probability at least $\delta_{\ref{lem:hypergraph-far}}(k)$, such a hypergraph is $\epsilon$-far from being simplex free and the failure probability of $\caA$ is at most $\delta_{\ref{lem:hypergraph-far}}(k)/2$, the probability that $\caA$ terminates before viewing a simplex must be less than $1-\delta_{\ref{lem:hypergraph-far}}(k)/2$.
Hence, Theorem~\ref{the:hypergraph-simplex-freeness} follows from the following lemma:
\begin{lemma}\label{lem:hypergraph-simplex-freeness}
  For any fixed $k$, any deterministic algorithm that detects a simplex in a hypergraph selected according to $\caD_{n,k}$ with probability at least $\delta_{\ref{lem:hypergraph-far}}(k)/2$ (over the choice of the hypergraph and internal randomness) requires $\Omega(n^{1/k})$ queries.
\end{lemma}

\subsubsection{Proof of Theorem~\ref{the:simplex-freeness}}\label{subsec:simplex-freeness}
  We first show that testing $\bfA$-freeness requires $\Omega(n^{1/k})$ queries.
  Suppose $\bfA $ is of the form
  \[
    (A = \set{a_1,\ldots,a_{k+1}},\set{R^\bfA}_{R \in \sigma}).
  \]
  From the assumption, for every $i \in [k+1]$, there exists a $k$-ary relation $R^\bfA = \set{(a_1,\ldots,a_{i-1},a_{i+1},\ldots,a_{k+1})}$.

  We reduce testing simplex-freeness on regular $k$-uniform hypergraphs to testing $\bfA$-freeness.
  We only consider hypergraphs generated by the distribution given in Section~\ref{subsubsec:simplices-distribution}.
  In particular, the input hypergraph $G=(V,\caE)$ is $d$-regular for $d = \Theta(n^{k-1-1/k})$ and the vertex set can be partitioned into $V_1,\ldots,V_{k+1}$ of the same size.
  For a set $I \subseteq [k+1]$ of size $k$, we define $\caE_I$ as the set of hyperedges spanning $V_i$'s for $i \in I$.

  Now, we construct a multiset relational structure $\bfB$ on the domain $V$ as follows.
  For a $k$-ary relation $R^\bfA$ of the form
  \[
  \set{(a_1,a_2,\ldots,a_{i-1},a_{i+1},\ldots,a_{k+1})},
  \] we define $R^\bfB = \caE_{[k+1]\setminus \set{i}}$.
  Here, we identify a tuple
  \[
    (v_1,\ldots,v_{i-1},v_{i+1},\ldots,v_{k+1})
  \]
  and a hyperedge
  \[
    \set{v_1,\ldots,v_{i-1},v_{i+1},\ldots,v_{k+1}}
  \] with $v_1 \in V_1,\ldots,v_{i-1} \in V_{i-1},v_{i+1} \in V_{i+1},\ldots,v_{k+1} \in V_{k+1}$.
  For any other relation $R^\bfA = \set{(a_{i_1},\ldots,a_{i_\ell})}$ of arity $\ell \leq k-1$, we define $R^\bfB$ as the set consisting of $|\caE| / \bar{n}^{\ell}$ many copies of each tuple in $V_{i_1} \times \cdots \times V_{i_\ell}$, where $\bar{n} = n/(k+1)$.
  For simplicity, we assume that $\bar{n}^{\ell}$ divides $|\caE|$ for every $\ell \leq k-1$, but we can easily extend the argument to the general case.
  Now, note that every relation in $\bfB$ has exactly $|\caE|$ many tuples, which means that $\|\bfB\| \leq \|\bfA\|\cdot |\caE|$

  Now, we check the reduction above satisfies the conditions of Definition~\ref{def:reduction}.
  If $G$ is simplex-free, then $\bfB$ is $\bfA$-free, which confirms (i).
  Suppose that $G$ is $\epsilon$-far from simplex-freeness, and $\bfB$ is $\delta$-close to simplex-freeness.
  Then, removing at most $\delta \|\bfB\|$ tuples from $\bfB$ corresponds to removing $O(\delta \|\bfA\|\|\bfB\|) = O(\delta \|\bfA\|^2 |\caE|)$ hyperedges from $G$.
  Hence, we have $\delta = \Omega(\epsilon /\|\bfA\|^2)$, which confirms (ii).
  It is easy to see that we can answer to a query to the completion oracle for $\bfB$ with a constant number of queries to the completion oracle for $G$, which confirms (iii).
  Hence, testing $\bfA$-freeness requires $\Omega(n^{1/k})$ queries by Theorem~\ref{the:hypergraph-simplex-freeness}.
  This concludes the proof of Theorem~\ref{the:simplex-freeness}.

\section{Unions of Conjunctive Queries}\label{sect:ucqs}

In this section, we give a proof of Corollary~\ref{cor:union-of-cqs}.

By the results in~\cite[Section 3]{BovaChen14-width-ep},
$\phi$ may be computably rewritten to the form
$\phi' = \phi_1 \vee \cdots \vee \phi_k$
that is normalized in the sense that
for distinct $i, j \in [k]$,
there does not exist a homomorphism from
$\bfA[\phi_i]$
to
$\bfA[\phi_j]$.
Note that, given a structure $\bfB$,
it holds that
$\phi$ evaluates to true on $\bfB$ if and only if
there exists $i \in [k]$ such that
there is a homomorphism from $\bfA[\phi_i]$ to $\bfB$
(equivalently, $\bfB$ is not $\bfA[\phi_i]$-free).

For the positive result, assume that
the condition in the statement of Corollary~\ref{cor:union-of-cqs}
holds.
Given a structure $\bfB$, use the positive algorithmic result of Theorem~\ref{the:main} with an error parameter $\epsilon/k$ to test, for each $i \in [k]$, whether or not $\bfB$ is $\bfA[\phi_i]$-free.
Reject to claim that $\phi$ is true if and only if
one of the tests rejects
(indicating that there is a homomorphism from some $\bfA[\phi_i]$ to $\bfB$).
This algorithm rejects with probability at least $2/3$ when $\bfB$ is $\epsilon$-far from $\phi$-falsity because then $\bfB$ is $\epsilon/k$-far from being $\bfA[\phi_i]$-free for some $i \in [k]$.

For the lower bound, assume that
the condition in the statement of Corollary~\ref{cor:union-of-cqs}
does not hold;
then, there exists $\ell \in [k]$
such that the core of $\bfA[\phi_\ell]$ is $\alpha$-acyclic.
By Theorem~\ref{the:main},
it suffices to reduce from testing $\bfA[\phi_\ell]$-freeness
to testing $\phi$-falsity.  We perform this by using the natural generalization of
Definition~\ref{def:reduction}.
The reduction is to map a multiset structure $\bfB$
to the product $\bfA[\phi_\ell] \times \bfB$,
by which we mean the structure with domain $A[\phi_\ell] \times B$
and where, for each relation symbol $R$, it holds that
$R^{\bfA[\phi_\ell] \times \bfB}
= \multiset{((a_1,b_1),\ldots,(a_r,b_r)) \mid
(a_1,\ldots,a_r) \in R^{\bfA[\phi_\ell]},
(b_1,\ldots,b_r) \in R^{\bfB}
}$; here, $r$ denotes the arity of $R$.
It is straightforward to verify that the first and second projections
are homomorphisms from
$\bfA[\phi_\ell] \times \bfB$
to
$\bfA[\phi_\ell]$ and  $\bfB$, respectively.
Set $\bfC = \bfA[\phi_\ell] \times \bfB$.

Let us verify that this is indeed a reduction.

(i) We claim that if $\bfB$ is $\bfA[\phi_\ell]$-free,
then $\bfA[\phi_\ell] \times \bfB$ causes $\phi$ to evaluate to false.
Let us argue the contrapositive.  Suppose that $\phi$ is true on
the product.  Then, there exists $i \in [k]$ such that
$\bfA[\phi_i]$ has a homomorphism to the product
$\bfC = \bfA[\phi_\ell] \times \bfB$; but since this product
has a homomorphism to $\bfA[\phi_\ell]$, it follows that
$\bfA[\phi_i]$ has a homomorphism to $\bfA[\phi_\ell]$,
implying (by normalization) that $i = \ell$.
Thus, $\bfA[\phi_\ell]$ has a homomorphism to this product,
and hence to $\bfB$.

(ii) Let $R$ be a relation symbol.
By the definition of the product $\bfA[\phi_\ell] \times \bfB$,
each tuple $(b_1, \ldots, b_r)$ in $R^{\bfB}$
gives rise to $\lvert R^{\bfA[\phi_\ell]}\rvert$ tuples in
$R^{\bfA[\phi_\ell] \times \bfB}$; let us call these the \emph{offshoots}
of the tuple $(b_1, \ldots, b_r)$. It is evident that each tuple in
$R^{\bfA[\phi_\ell] \times \bfB}$ is the offshoot of exactly one tuple
in $R^{\bfB}$.
Let $K$ be the maximum, over relation symbols $R$, of $|R^{\bfA[\phi_\ell]}|$.
We have $||\bfC|| \leq K ||\bfB||$.

We claim that if a structure $\bfB$ is $\epsilon$-far from being
$\bfA[\phi_\ell]$-free,
then $\bfC$ is $\epsilon/K$-far from $\phi$-falsity.
Suppose that $\bfC'$ is obtained from $\bfC$ by removing $(\epsilon/K) ||\bfC||$
tuples from relations.
Let us say that a tuple $(b_1, \ldots, b_r)$ in a relation $R^{\bfB}$ of $\bfB$
is \emph{partly removed in $\bfC'$} if one of its offshoots was removed
to obtain $\bfC'$ from $\bfC$.
Then, the number of tuples of $\bfB$ that are partly removed in $\bfC'$
is bounded above by
$(\epsilon/K) ||\bfC|| \leq (\epsilon/K) K ||\bfB|| = \epsilon ||\bfB||$.
Let $\bfB'$ be the structure obtained from $\bfB$ by
removing each tuple that is partly removed in $\bfC'$.
By assumption, there exists a homomorphism $g$ from
$\bfA[\phi_\ell]$
to $\bfB'$.
It is straightforward to verify that the mapping
$h$ defined by $h(a) = (a,g(a))$ is a homomorphism from
$\bfA[\phi_\ell]$ to $\bfC'$,
and so $\phi$ is true on $\bfC'$.

(iii) Consider a query $(R, (c_1, \ldots, c_r))$
concerning the relation $R^{\bfC}$; here,
$(c_1, \ldots, c_r) \in {(C \cup \{ * \})}^r$
Define $(a_1, \ldots, a_r) \in {(A[\phi_\ell] \cup \{ * \})}^r$
to be the tuple such that $a_i = *$ if $c_i = *$,
and $a_i = d_i$ if $c_i$ has the form $(d_i,e_i)$.
Analogously,
define $(b_1, \ldots, b_r) \in {(B \cup \{ * \})}^r$
to be the tuple such that $b_i = *$ if $c_i = *$,
and $b_i = e_i$ if $c_i$ has the form $(d_i,e_i)$.
If the query is a completion query, then to answer it,
one can choose a random completion $\bfa'$ of $(a_1, \ldots, a_r)$,
call the completion query for $\bfB$ to obtain a random completion $\bfb'$
of $(b_1, \ldots, b_r)$, and then return the product of $\bfa'$ and $\bfb'$.
If the query is a size query, then to answer it,
one can count the number $M$ of completions of $(a_1, \ldots, a_r)$,
call the size query for $\bfB$ to obtain the number $N$ of completions of
$(b_1, \ldots, b_r)$,
and return $M \cdot N$.

(iv) We have $|C| = |A[\phi_\ell]| \cdot |B|$.

\paragraph*{Acknowledgements.} The first author thanks
Am\'elie Gheerbrant  and Cristina Sirangelo for a useful discussion.

\bibliographystyle{ACM-Reference-Format}
\bibliography{main}

@inproceedings{Adler:2018jv,
author = {Adler, Isolde and Harwath, Frederik},
title = {{Property Testing for Bounded Degree Databases}},
booktitle = {STACS},
year = {2018},
pages = {6:1--6:14},
}

@article{Bentley:1979cv,
author = {Bentley, Jon Louis},
title = {{Decomposable searching problems}},
journal = {Information Processing Letters},
year = {1979},
volume = {8},
number = {5},
pages = {244--251},
}

@inproceedings{BenMoshe11,
  author    = {Sagi Ben{-}Moshe and
               Yaron Kanza and
               Eldar Fischer and
               Arie Matsliah and
               Mani Fischer and
               Carl Staelin},
  title     = {Detecting and exploiting near-sortedness for efficient relational
               query evaluation},
  booktitle = {ICDT},
  pages     = {256--267},
  year      = {2011},
  }

@article{Fagin:2005kn,
author = {Fagin, Ronald and Kolaitis, Phokion G and Miller, Ren{\'e}e J and Popa, Lucian},
title = {Data exchange: semantics and query answering},
journal = {Theoretical Computer Science},
year = {2005},
volume = {336},
number = {1},
pages = {89--124},
}

@inproceedings{BovaChen14-width-ep,
 author    = {Simone Bova and
               Hubie Chen},
  title     = {The Complexity of Width Minimization for Existential Positive
               Queries},
  booktitle = {ICDT},
  year      = {2014},
  pages     = {235-244},
}

@article{embedding,
  author    = {Hubie Chen and
               Moritz M{\"{u}}ller},
  title     = {The Parameterized Space Complexity of Embedding Along a Path},
  journal   = {Theory Comput. Syst.},
  volume    = {61},
  number    = {3},
  pages     = {851--870},
  year      = {2017},
}

@inproceedings{Joos:2017tb,
author = {Joos, Felix and Kim, Jaehoon and K{\"u}hn, Daniela and Osthus, Deryk},
title = {A characterization of testable hypergraph properties},
booktitle = {FOCS},
year = {2017},
pages = {859--867},
}

@article{Fox:2016bp,
author = {Fox, Jacob and Pach, J{\'a}nos and Suk, Andrew},
title = {A Polynomial Regularity Lemma for Semialgebraic Hypergraphs and Its Applications in Geometry and Property Testing},
journal = {SIAM Journal on Computing},
year = {2016},
volume = {45},
number = {6},
pages = {2199--2223},
}

@inproceedings{Rodl:2007dt,
author = {R{\"o}dl, V and Schacht, M},
title = {Property testing in hypergraphs and the removal lemma},
booktitle = {STOC},
year = {2007},
pages = {488--495},
}

@article{ron2010algorithmic,
  Author =       {Dana Ron},
  Title =        {Algorithmic and Analysis Techniques in Property
                  Testing},
  Journal =      {Foundations and Trends in Theoretical Computer Science},
  Volume =       5,
  Issue =        2,
  Year =         2010,
  Pages =        {73--205},
}

@article{rubinfeld2011survey,
  Author =       {Ronitt Rubinfeld and Asaf Shapira},
  Journal =      {Electronic Colloquium on Computational Complexity
                  (ECCC)},
  Title =        {Sublinear Time Algorithms},
  Volume =       18,
  note =         {TR11-013},
  Year =         2011
}

@book{goldreich2010book,
  Editor =       {Oded Goldreich},
  Title =        {Property Testing: Current Research and Surveys},
  Publisher =    {Springer},
  Series =       {LNCS},
  Volume =       6390,
  Year =         2010
}

@article{HellNesetril92-core,
author = {P. Hell and J. Nesetril},
title = {The core of a graph},
journal = {Discrete Mathematics},
volume = {109},
year = {1992},
pages = {117-126}}

@article{Chen14-existentialpositive,
  author    = {Hubie Chen},
  title     = {On the complexity of existential positive queries},
  journal   = {ACM Trans. Comput. Log.},
  volume    = {15},
  number    = {1},
  year      = {2014},
}

@article{ChenLarose17-metaquestions,
  author    = {Hubie Chen and
               Benoit Larose},
  title     = {Asking the Metaquestions in Constraint Tractability},
  journal   = {{TOCT}},
  volume    = {9},
  number    = {3},
  pages     = {11:1--11:27},
  year      = {2017},
  }

@article{ChenMueller15-fineclass,
  author    = {Hubie Chen and
               Moritz M{\"{u}}ller},
  title     = {The Fine Classification of Conjunctive Queries and Parameterized Logarithmic
               Space},
  journal   = {Transactions on Computation Theory},
  volume    = {7},
  number    = {2},
  pages     = {7},
  year      = {2015},
}

@article{ChenMueller17-hierarchy,
  author    = {Hubie Chen and
               Moritz M{\"{u}}ller},
  title     = {One Hierarchy Spawns Another: Graph Deconstructions and the Complexity
               Classification of Conjunctive Queries},
  journal   = {{ACM} Trans. Comput. Log.},
  volume    = {18},
  number    = {4},
  pages     = {29:1--29:37},
  year      = {2017},
}

@inProceedings{BartoKozik09-boundedwidth,
author = {L. Barto and M. Kozik},
title = {Constraint Satisfaction Problems of Bounded Width},
booktitle = {FOCS},
pages = {595-603},
year = 2009}

@inproceedings{Marx10-tractablehypergraph,
  author    = {D{\'a}niel Marx},
  title     = {Tractable hypergraph properties for constraint satisfaction and conjunctive queries},
  booktitle ="STOC",
  year="2010",
  pages="735-744",
}

@Article{Grohe07-otherside,
  title = {The complexity of homomorphism and constraint satisfaction problems seen from the other side},
    author = {Martin Grohe},
journal = {Journal of the ACM},
number = {1},
volume = {54},
year = {2007}}

@article{GottlobLeoneScarcello01-complexityacyclic,
  author = {Georg Gottlob and Nicola Leone and Francesco Scarcello},
  year = {2001},
  title = {The complexity of acyclic conjunctive queries},
  journal = {Journal of the ACM},
  volume = {48},
  number = {3},
  pages = {431-498}
}

@article{GottlobLeoneScarcello02-hypertreedecomposisions,
  author = {Georg Gottlob and Nicola Leone and Francesco Scarcello},
  year = {2002},
  title = {Hypertree decompositions and tractable queries},
  journal = {Journal of Computer and System Sciences},
  volume = {64},
  number = {3},
  pages = {579--627}
}

@article{PapadimitriouYannakakis99-database,
  author = {C. Papadimitriou and M. Yannakakis},
  title = {{On the Complexity of Database Queries}},
  journal = {Journal of Computer and System Sciences},
  volume = {58},
  year = {1999},
  number = {3},
  pages = {407-427}
}

@book{AbiteboulHullVianu95-foundationsdatabases,
  author = {S. Abiteboul and R. Hull and V. Vianu},
  title = {Foundations of Databases},
  publisher = {Addison-Wesley},
  year = {1995}
}

@Article{FederVardi99-structure,
  author =   {T. Feder and M. Vardi},
  title =    {The computational structure of monotone monadic {SNP} and constraint satisfaction: {A} study through {D}atalog and group theory},
  journal =    {SIAM Journal on Computing},
  year =   {1999},
  volume =   {28},
  pages =  {57--104},
}

@incollection{SchweikardtSchwentickSegoufin09-querylanguages,
  title = "Database Theory: Query Languages",
  author = "Nicole Schweikardt and Thomas Schwentick and Luc Segoufin",
  chapter = "19",
  booktitle =  "Algorithms and Theory of Computation Handbook",
  editor = "Mikhail J. Atallah and Marina Blanton",
  publisher = "CRC Press",
  edition = "Second",
  volume = "2: Special Topics and Techniques",
  year = "2009",
}

@inproceedings{ChaudhuriV93,
  author    = {Surajit Chaudhuri and
               Moshe Y. Vardi},
  title     = {Optimization of \emph{Real} Conjunctive Queries},
  booktitle = {PODS},
  pages     = {59--70},
  year      = {1993},
  }

@inproceedings{Dasgupta:2014kg,
author = {Dasgupta, Anirban and Kumar, Ravi and Sarlos, Tamas},
title = {On estimating the average degree},
booktitle = {WWW},
year = {2014},
pages = {795--806},
}

@article{Goldreich:2008in,
author = {Goldreich, Oded and Ron, Dana},
title = {Approximating average parameters of graphs},
journal = {Random Structures {\&} Algorithms},
year = {2008},
volume = {32},
number = {4},
pages = {473--493}
}

@inproceedings{Chandra:1977ez,
author = {Chandra, Ashok K and Merlin, Philip M},
title = {Optimal implementation of conjunctive queries in relational data bases},
booktitle = {STOC},
year = {1977},
pages = {77--90},
}

@article{BB16,
  author    = {Johann Brault{-}Baron},
  title     = {Hypergraph Acyclicity Revisited},
  journal   = {ACM Computing Surveys},
  volume    = {49},
  number    = {3},
  pages     = {54:1--54:26},
  year      = {2016},
}

@inproceedings{Yoshida:2011da,
author = {Yoshida, Yuichi},
title = {Optimal constant-time approximation algorithms and (unconditional) inapproximability results for every bounded-degree {CSP}},
booktitle = {STOC},
year = {2011},
pages = {665--674},
}

@inproceedings{Sohler:2012kqa,
author = {Sohler, Christian},
title = {Almost Optimal Canonical Property Testers for Satisfiability},
booktitle = {FOCS},
year = {2012},
pages = {541--550},
}

@inproceedings{Alon:2002ua,
author = {Alon, Noga and Shapira, Asaf},
title = {Testing satisfiability},
booktitle = {SODA},
year = {2002},
pages = {645--654},
}

@article{Kaufman:2004vg,
author = {Kaufman, Tali and Krivelevich, Michael and Ron, Dana},
title = {Tight Bounds for Testing Bipartiteness in General Graphs},
journal = {SIAM Journal on Computing},
year = {2004},
volume = {33},
number = {6},
pages = {1441--1483},
}

@article{Rubinfeld:1996um,
author = {Rubinfeld, Ronitt and Sudan, Madhu},
title = {Robust Characterizations of Polynomials with Applications to Program Testing},
journal = {SIAM Journal on Computing},
year = {1996},
volume = {25},
number = {2},
pages = {252--271},
}

@article{Goldreich:2002bn,
author = {Goldreich, Oded and Ron, Dana},
title = {Property Testing in Bounded Degree Graphs},
journal = {Algorithmica},
year = {2002},
volume = {32},
number = {2},
pages = {302--343},
}

@article{Alon:2003jp,
author = {Alon, Noga and de la Vega, W Fernandez and Kannan, Ravi and Karpinski, Marek},
title = {Random sampling and approximation of MAX-CSPs},
journal = {Journal of Computer and System Sciences},
year = {2003},
volume = {67},
number = {2},
pages = {212--243},
}

@article{Goldreich:1998wa,
author = {Goldreich, Oded and Goldwasser, Shafi and Ron, Dana},
title = {Property testing and its connection to learning and approximation},
journal = {Journal of the ACM},
year = {1998},
volume = {45},
number = {4},
pages = {653--750}
}

@article{Blum:1993cn,
author = {Blum, Manuel and Luby, Michael and Rubinfeld, Ronitt},
title = {Self-testing/correcting with applications to numerical problems},
journal = {Journal of Computer and System Sciences},
year = {1993},
volume = {47},
number = {3},
pages = {549--595},
}

@article{BenSasson:2005we,
author = {Ben-Sasson, Eli and Harsha, Prahladh and Raskhodnikova, Sofya},
title = {Some {3CNF} Properties Are Hard to Test},
journal = {SIAM Journal on Computing},
year = {2005},
volume = {35},
number = {1},
pages={1--21}
}

@inproceedings{Fischer:2002ev,
author = {Fischer, Eldar and Lehman, Eric and Newman, Ilan and Raskhodnikova, Sofya and Rubinfeld, Ronitt and Samorodnitsky, Alex},
title = {Monotonicity testing over general poset domains},
booktitle = {STOC},
year = {2002},
pages = {474--483},
}

@article{Yoshida:2014zn,
author = {Yoshida, Yuichi},
title = {Testing list {$H$}-homomorphisms},
journal = {Computional Complexity},
year = {2016},
volume = {25},
number = {4},
pages = {737--773}
}

@article{Alon:2008gn,
author = {Alon, Noga and Kaufman, Tali and Krivelevich, Michael and Ron, Dana},
title = {Testing Triangle-Freeness in General Graphs},
journal = {SIAM Journal on Discrete Mathematics},
year = {2008},
volume = {22},
number = {2},
pages = {786--819},
}

@article{Alon:2009gn,
author = {Alon, Noga and Fischer, Eldar and Newman, Ilan and Shapira, Asaf},
title = {A Combinatorial Characterization of the Testable Graph Properties: It's All About Regularity},
journal = {SIAM Journal on Computing},
year = {2009},
volume = {39},
number = {1},
pages = {143--167},
}

@inproceedings{Yoshida:2014tq,
author = {Yoshida, Yuichi},
title = {A Characterization of Locally Testable Affine-Invariant Properties via Decomposition Theorems},
booktitle = {STOC},
year = {2014},
pages = {154--163}
}

@inproceedings{Yoshida:2016zz,
author = {Yoshida, Yuichi},
title = {Gowers Norm, Function Limits, and Parameter Estimation},
booktitle = {SODA},
year = {2014},
pages = {1391--1406}
}

@inproceedings{Bhattacharyya:2013fa,
author = {Bhattacharyya, Arnab and Yoshida, Yuichi},
title = {An Algebraic Characterization of Testable Boolean {CSPs}},
booktitle = {ICALP},
year = {2013},
pages = {123--134},
}

@inproceedings{Chen:2016wf,
author = {Chen, Hubie and Valeriote, Matt and Yoshida, Yuichi},
title = {Testing Assignments to Constraint Satisfaction Problems},
booktitle = {FOCS},
year = {2016},
pages = {525--534}
}

@article{Iwama:2018bs,
author = {Iwama, Kazuo and Yoshida, Yuichi},
title = {Parameterized Testability},
journal = {ACM Transactions on Computation Theory},
year = {2018},
volume = {9},
number = {4},
pages = {16--16},
}

@inproceedings{Zhuk17-dichotomy,
  author    = {Dmitriy Zhuk},
  title     = {A Proof of {CSP} Dichotomy Conjecture},
  booktitle = {FOCS},
  pages     = {331--342},
  year      = {2017},
  }

@inproceedings{Bulatov17-dichotomy,
  author    = {Andrei A. Bulatov},
  title     = {A Dichotomy Theorem for Nonuniform CSPs},
  booktitle = {FOCS},
  pages     = {319--330},
  year      = {2017},
  }

\appendix

\section{Proof of Theorem~\ref{thm:reduction-symbol-distinct}}

\begin{proof}
Let $\bfA = (A,{\{ R^{\bfA} \}}_{R \in \sigma})$
be a core.  We show how to construct
a symbol-distinct structure $\bfA'$ such that
testing $\bfA'$ reduces to testing $\bfA$.

Define $\sigma'$
as the vocabulary $\set{ R_{\bft} \mid R \in \sigma, \bft \in R^{\bfA} }$,
where the arity of $R_{\bft}$ is equal to that of $R \in \sigma$.
Define $\bfA'$ to be the structure
$(A, {\{ R_{\bft} \}}_{R_{\bft} \in \sigma'})$
where $R_{\bft}^{\bfA'} = \{ \bft \}$.
For any structure $\bfB'$ with domain $B'$ and vocabulary $\sigma'$,
we construct a structure $\bfB$ in the following way.
The domain of $\bfB$ is $A \times B'$,
and for each $R \in \sigma$,
we define
$R^{\bfB} = \multiset{ ((a_1,b_1),\ldots, (a_r,b_r)) \mid
(a_1, \ldots, a_r) \in R^{\bfA},
(b_1, \ldots, b_r) \in R_{(a_1, \ldots, a_r)}^{\bfB'} }$.
We have $\| \bfB \| = \| \bfB' \|$; moreover,
there is clearly a natural bijection between the tuples
of $\bfB'$ and the tuples of $\bfB$:
namely, map the tuple
$(b_1, \ldots, b_r) \in R_{(a_1, \ldots, a_r)}^{\bfB'}$
to the tuple
$((a_1,b_1),\ldots, (a_r,b_r)) \in R^{\bfB}$.

Let $\bfB_-$ be a structure obtained from $\bfB$ by deleting tuples,
and let $\bfB'_-$ be the corresponding structure obtained from $\bfB'$
by deleting the corresponding tuples under the bijection just given.
We prove that $\bfB_-$ is $\bfA$-free if and only if
$\bfB'_-$ is $\bfA'$-free, yielding properties (i) and (ii)
in the definition of reduction.
For any homomorphism $h$ from $\bfA$ to $\bfB_-$,
the composition $\pi_1(h)$ is readily seen to be an endomorphism
of $\bfA$.  (Here, $\pi_1$ denotes the first projection.)
Since each endomorphism of $\bfA$ is a bijection
(as $\bfA$ is a core),
when $h$ is a homomorphism from $\bfA$ to $\bfB_-$,
by composing $h$ with an appropriate power of $\pi_1(h)$,
we obtain a homomorphism $g$ from $\bfA$ to $\bfB_-$
such that $\pi_1(g)$ is the identity mapping.
Hence, if there is any homomorphism from $\bfA$ to $\bfB_-$
at all, there is a homomorphism $g$ such that
$\pi_1(g)$ is the identity mapping; moreover,
when such a homomorphism exists, the image $g(A)$ of $g$
is straightforwardly verified to be
the graph of a homomorphism from $\bfA'$ to $\bfB'_-$.
In the other direction, if there is a homomorphism $f$
from $\bfA'$ to $\bfB'_-$, then the mapping
$a \mapsto (a,f(a))$ is readily verified to be a homomorphism
from $\bfA$ to $\bfB_-$.

We verify property~(iii) in the definition of reduction,
as follows.
Consider a query $(R, (q_1, \ldots, q_r))$ for $\bfB$.
\begin{itemize}

\item 
Let $(c_1, \ldots, c_r) \in {(A \cup \{ * \})}^r$
be the tuple such that
$q_i = *$ implies $c_i = *$ and
$q_i \neq *$ implies $c_i = \pi_1(q_i)$.

\item 
Let $(d_1, \ldots, d_r) \in {(A \cup \{ * \})}^r$
be the tuple such that
$q_i = *$ implies $d_i = *$ and
$q_i \neq *$ implies $d_i = \pi_2(q_i)$.
\end{itemize}
Let $\bft_1, \ldots, \bft_k$ be the tuples
in  $R^{\bfA}$ that are
completions of $(c_1, \ldots, c_r)$.
We may use a size query
$(R_{\bft_i}, (d_1, \ldots, d_r))$ for $\bfB'$ to determine
the number of completions $m_i$ of $(d_1, \ldots, d_r)$ in
$R_{\bft_i}^{\bfB'}$.
If the query $(R, (q_1, \ldots, q_r))$
was a size query, then the answer
straightforwardly verified, from the definition of $\bfB$,
to be $m_1 + \cdots + m_k$.
If the query $(R, (q_1, \ldots, q_r))$
was a completion query,
then it is answered in the following way.
If $m_1 + \cdots + m_k = 0$, then return $\bot$.
Otherwise, randomly pick an index $i \in [k]$,
by selecting the index $i \in [k]$ with probability
$m_i / (m_1 + \cdots + m_k)$;
return the answer to a completion query
$(R_{\bft_i}, (d_1,\ldots,d_r))$ for $\bfB'$. Correctness is verified, again,
from the definition of $\bfB$.

Property~(iv) is clear: $|B| = |A| \cdot |B'|$.
\end{proof}

\section{Proof of
Theorem~\ref{thm:reduction-induced-hypergraph}}

\begin{proof}
Let $\sigma'$ be the vocabulary
and let $\bfA'$ be the symbol-distinct structure
with domain $S$
that are
constructed as follows.
For each $R \in \sigma$,
if the unique tuple $(a_1, \ldots, a_r) \in R^{\bfA}$
contains at least one entry in $S$,
then let $i_1 <  \cdots < i_p$ be the indices from $[r]$
such that $a_{i_1}, \ldots, a_{i_p}$ are the entries
among $a_1, \ldots, a_r$ that are contained in $S$;
and, include a symbol $R'$ in $\sigma'$
whose arity is $p$, where $R'^{\bfA'} = \{ (a_{i_1}, \ldots, a_{i_p}) \}$.
Let $\bfB'$ be any structure with domain $B'$ over $\sigma'$.
From $\bfB'$, we define a structure $\bfB$ as follows.
The domain of $\bfB$ is $B = B' \cup \{ + \}$.
For each $R' \in \sigma'$, let $i_1 < \cdots < i_p$
be the indices defined above,
and
define $R^{\bfB}$
as the relation that contains,
for each tuple $(b'_1, \ldots, b'_p) \in R^{\bfB'}$,
the tuple $(b_1, \ldots, b_r)$
where $b_{i_1} = b'_1, \ldots, b_{i_p} = b'_p$,
and where all other entries are equal to $+$.
Clearly, there is a natural bijection between tuples in
$R^{\bfB'}$ and $R^{\bfB}$.
For each symbol $R \in \sigma$ that does not have a corresponding
symbol $R' \in \sigma'$,
define $R^{\bfB}$ as $\multiset{ (+, \ldots, +) }$,
where the multiplicity of $(+, \ldots, +)$ is $\| \bfB'\|$;
let $Z$ denote the number of such symbols $R \in \sigma$,
and call each such symbol \emph{trivialized}.
We have $\| \bfB \| = (Z+1) \| \bfB'\|$.

It is straightforward to establish that
a mapping $f: S \to B'$ is a homomorphism
from $\bfA'$ to $\bfB'$
if and only if the \emph{canonical} extension $f': A \to B$
of $f$ that sends each element in $A \setminus S$ to $+$
is a homomorphism from $\bfA$ to $\bfB$.
This establishes property~(i) in the definition of reduction.

Let us establish property~(ii) in the definition of reduction.
We claim that if $\bfB'$ is $\epsilon'$-far from being $\bfA'$-free,
then $\bfB$ is $\epsilon' / (Z+1)$-far from being $\bfA$-free.
Consider the deletion of $(\epsilon' / (Z+1)) \| \bfB \| = \epsilon' \| \bfB' \|$
tuples from $\bfB$.
Since each trivialized symbol $R$ contains $\|\bfB'\|$ copies of
$(+, \ldots, +)$, in effect, the deletion amounts to
the deletion of at most $\epsilon' \| \bfB' \|$ tuples from
the relations $R^{\bfB}$, over non-trivialized symbols $R$.
Under the bijection given above, this is equivalent
to the deletion of at most $\epsilon' \| \bfB' \|$ tuples from
the relations of $\bfB'$.  Since $\bfB'$ is assumed to be
$\epsilon'$-far from being $\bfA'$-free,
this deletion of tuples from $\bfB'$ results in a structure
that admits a homomorphism from $\bfA'$.
By taking the canonical extension of this homomorphism,
we obtain a homomorphism from $\bfA$ to $\bfB$ under the deletions,
establishing our claim.

Let us establish property~(iii) in the definition of reduction.
A size query for $\bfB$ can clearly be resolved using
size queries to $\bfB'$:
for each trivialized symbol $R$, the size of $R^{\bfB}$ is $\| \bfB' \|$,
and for each other symbol $R$, the size of $R^{\bfB}$ is equal
to that of $R^{\bfB'}$.
A completion query $(R, \bfq)$
for $\bfB$ can be resolved in the following way.
If the query concerns a trivialized symbol $R$,
then the answer is $(+, \ldots, +)$ if this tuple is a completion of
$\bfq$, and $\bot$ otherwise.
Otherwise, an answer can be obtained from
a completion query $(R, \bfq')$ for $\bfB'$.

Property~(iv) is clear: $|B| = |B'| + 1$.
\end{proof}

\section{Missing Proofs for Section~\ref{subsec:simplices}}\label{sec:simplices-appendix}
\begin{proof}[Proof of Lemma~\ref{lem:hypergraph-far}]
  We shall show that with high probability over the choice of a hypergraph $G=(V,\caE)$ according to $\caD_{n,k}$, there are at least $\gamma |\caE|$ edge-disjoint simplices in the hypergraph, for some constant $\gamma$.

  Let the vertices in each set $V_b\;(b \in [k+1])$ be denoted by $v_{b,1},\ldots,v_{b,\bar{n}}$.
  For a tuple $\bfb = (b_1,\ldots,b_k)\in {[k+1]}^{k}$ of distinct indices and a tuple $\bfi = (i_1,\ldots,i_k) \in {[\bar{n}]}^k$, let $\alpha^{\bfb}_{\bfi}$ be an indicator of the event that there is a hyperedge spanning $v_{b_1,i_1},\ldots,v_{b_k,i_k}$.
  Then, $\Pr[\alpha^\bfb_\bfi=1] = 1 - {\left( 1 - \frac{1}{\bar{n}^{k-1}} \right)}^{\bar{d}}$ and so $\Pr[\alpha^\bfb_\bfi=1] \leq \frac{\bar{d}}{\bar{n}^{k-1}}$, and
  \[
    \Pr[\alpha^\bfb_\bfi=1]
    \geq
    \frac{\bar{d}}{\bar{n}^{k-1}} - {\bar{d} \choose 2}\frac{1}{\bar{n}^{2(k-1)}}
    \geq \frac{\bar{d}}{\bar{n}^{k-1}} \left( 1- \frac{1}{ 2(k+1) \bar{n}^{1/k}}\right)
  \]

  Next, consider any choice of $\bfi = (i_1,\ldots,i_{k+1}) \in {[\bar{n}]}^{k+1}$.
  Let $X_\bfi$ denote the indicator of the event that there is a simplex spanning the vertices $v_{1,i_1},\ldots,v_{k+1,i_{k+1}}$.
  Since the choices of the $k$-dimensional matchings on $V_{b_1},\ldots,V_{b_k}$ for $k$-sized subsets $\set{b_1,\ldots,b_k}$ of $[k+1]$ are mutually independent, $\Pr[X_\bfi=1] \leq {\left(\frac{\bar{d}}{\bar{n}^{k-1}}\right)}^{k+1}$ holds, and
  \[
    \Pr[X_\bfi=1]
    \geq
    {\left(\frac{\bar{d}}{\bar{n}^{k-1}}\right)}^{k+1} {\left( 1- \frac{1}{ 2(k+1) \bar{n}^{1/k}}\right)}^{k+1}.
  \]
  Let $\beta_\bfi$ denote the indicator of the event that there is a simplex different from $v_{1,i_1},\ldots,v_{k+1,i_{k+1}}$ that includes one of the hyperedges.
  Since the probability of having a hyperedge $(v_{b_1,i'_1},\ldots,v_{b_k,i'_k})$ conditioned on having the hyperedge $(v_{b_1,i_1},\ldots,v_{b_k,i_k})$ for any choice of 
  \[ |\set{i'_1,\ldots,i'_k} \cap \set{i_1,\ldots,i_k}| \geq 1 \]
  is at most $1 - {(1-\frac{1}{\bar{n}^{k-1}})}^{\bar{d}-1} \leq \frac{\bar{d}-1}{\bar{n}^{k-1}}$,
  \begin{align*}
    & \Pr[X_{\bfi} = 1 \wedge \beta_\bfi = 1]
    =
    \Pr[X_{\bfi} = 1] \cdot \Pr[\beta_\bfi = 1 \mid X_\bfi = 1] \\
    & \leq
    {\left(\frac{\bar{d}}{\bar{n}^{k-1}}\right)}^{k+1}
    \cdot
    (k+1)
    \cdot
    (\bar{n}-1)
    \cdot
    {\left(\frac{\bar{d}-1}{\bar{n}^{k-1}}\right)}^{k} \\
    & \leq
    {\left(\frac{\bar{d}}{\bar{n}^{k-1}}\right)}^{k+1}
    \frac{(k+1){(\bar{d}-1)}^k}{\bar{n}^{k^2-k-1}}
  \end{align*}
  Let $\eta_\bfi$ be the indicator of the event that there is a simplex spanning the vertices $v_{1,i_1},\ldots, v_{k+1,i_{k+1}}$ and there is no other simplex that shares any of its hyperedges.
  Then, we have
  \begin{align*}
    & \Pr[\eta_\bfi=1] =  \Pr[X_\bfi = 1 \wedge \beta_\bfi = 0] \\
    & =
    \Pr[X_\bfi = 1] - \Pr[X_\bfi = 1 \wedge \beta_\bfi = 0] \\
    & \geq
    {\left(\frac{\bar{d}}{\bar{n}^{k-1}}\right)}^{k+1}
    \cdot
    \left(
    {\left( 1- \frac{1}{ 2(k+1) \bar{n}^{1/k}}\right)}^{k+1} - \frac{(k+1){(\bar{d}-1)}^k}{\bar{n}^{k^2-k-1}}
    \right) \\
    & \geq \frac{1}{2}{ \left(\frac{\bar{d}}{\bar{n}^{k-1}}\right)}^{k+1}.
  \end{align*}
  under our assumption that $\bar{n}$ is sufficiently large.
  Hence, the expected number of tuples $(v_{1,i_1},\ldots,v_{k+1,i_{k+1}})$ that constitute a simplex that does not share a hyperedge with any other simplex is
  \[
    \sum_{\bfi}\eta_{\bfi} \geq
    \bar{n}^{k+1}
    \cdot \frac{1}{2}{\left(\frac{\bar{d}}{\bar{n}^{k-1}}\right)}^{k+1}
    =
    \frac{\bar{d}\bar{n}}{2{(k+1)}^k}
    =
    \frac{dn}{2k{(k+1)}^{k+1}}.
  \]

  Note that we have $\sum_{\bfi}\eta_\bfi \leq dn/k(k+1)$ as we only have $dn/k$ hyperedges and each simplex uses $k+1$ hyperedges.
  Let $\alpha = \frac{1}{4k{(k+1)}^{k+1}}$ and $p = \Pr[\sum_{\bfi}\eta_\bfi \leq \alpha dn]$.
  Then, we have
  \begin{align*}
    & \frac{dn}{2k{(k+1)}^{k+1}} \leq \E\left[\sum_{\bfi}\eta_\bfi\right]
    \leq p\alpha dn + (1 - p) \frac{dn}{k(k+1)} \\
  \end{align*}
The last quantity is bounded above by
$\alpha dn + (1-p)\frac{dn}{k(k+1)}$, implying
 $p \leq 1 - \frac{1}{4{(k+1)}^k}$.
\end{proof}

We will also use the following, which states that, for every set of $k-1$ vertices belonging to different blocks, the number of hyperedges containing them is high.
\begin{lemma}\label{lem:hypergraph-degree}
  For any fixed $k$, a hypergraph sampled uniformly from $\caD_{n,k}$ satisfies the following with probability $1-O(1/n^{2k})$:
  For any set $S$ of $k-1$ vertices belonging to different blocks and block $V_b$ not containing any vertex in $S$, the number $m_{S,b}$ of hyperedges of the form $S \cup \set{v}$ for $v \in V_b$ is $\Omega(n^{1-1/k})$.
\end{lemma}
\begin{proof}
  Let $S = \set{v_1,\ldots,v_{k-1}}$ and let $V_{b_1},\ldots,V_{b_{k-1}}$ be blocks to which $v_1,\ldots,v_{k-1}$ belong, respectively.
  For each $i \in [\bar{d}]$, the probability that the $i$-th $k$-dimensional matching spanning $V_b$ and $V_{b_1},\ldots,V_{b_{k-1}}$ has a hyperedge containing $S$ is $p:=1/\bar{n}^{k-2}$.
  Hence, $\mu := \E[m_{S,b}] = p \bar{d} = \bar{n}^{1-1/k}/(k+1)$.
  By Chernoff's inequality, $\Pr\left[m_{S,b} < \frac{\mu}{2} \right] \leq \exp\left(-\frac{\mu}{8} \right) \leq \frac{1}{10n^{2k}}$.
  Here in the last inequality, we use the fact that $n$ is sufficiently large.
  By union bound over the choice of $S$ and $b$, we have $m_{S,b} \geq \bar{n}^{1-1/k}/(2k+2)$ for every $S$ and $b$ with probability at least $1-O(1/n^{2k})$.
\end{proof}

\begin{proof}[Proof of Lemma~\ref{lem:hypergraph-simplex-freeness}]
  It will be convenient to view hypergraphs in the support of $\caD_{n,k}$ as being represented by $k$-dimensional matchings over tables.
  Namely, there is a table $T_{b,B}$ for each $b \in [k+1]$ and a set $B \ni b$ of size $k$.
  Each table $T_{b,B}$ is of size $\bar{n} \times \bar{d}$ (recall that $\bar{n} = n/(k+1)$ and $\bar{d}=d/k$): there is a row for each vertex $v$ in $V_b$, and each entry in $v$'s row corresponds to one of the $\bar{d}$ hyperedges that are incident to $v$ and to vertices in $V_{b'}$ with $b' \in B$.
  For $B = \set{b_1,\ldots,b_k}$, a hyperedge between $v_1 \in V_{b_1},\ldots,v_k \in V_{b_k}$ is represented by $T_{b_1,B}[v_1, i_1],\ldots,T_{b_k,B}[v_k, i_k]$ for some $i_1,\ldots,i_k \in [\bar{d}]$.
  Hence, the $\bar{d}$ perfect $k$-dimensional matchings among $V_{b_1},\ldots,V_{b_k}$ correspond to a single perfect $k$-dimensional matching among the entries of the tables $T_{b_1,B},\allowbreak T_{b_2,B},\ldots,T_{b_k,B}$, where $B= \set{b_1,\ldots,b_k}$.

  Let $\caA$ be a deterministic algorithm that performs $q(n)=o(n^{1/k})$ queries and whose goal is to detect a simplex with probability at least $\delta_{\ref{lem:hypergraph-far}}(k)$.
  The probability is taken over the choice of the hypergraph $G$ in the support of $\caD_{n,k}$ and internal randomness.
  Namely $\caA$ is a (possibly probabilistic) mapping from query-answer histories $\langle (q_1, a_1), \ldots , (q_t, a_t)\rangle$, to $q_{t+1}$ for every $t < q(n)$.
  We note that the mapping from query-answer histories to queries needs to be defined only on histories that are consistent with some hypergraph in the support of $\caD_{n,k}$.

  In what follows we define a process that answers the queries of the algorithm while generating a hypergraph according to $\caD_{n,k}$.
  At any time $t$, the queries of the algorithm and the answers to them determine the knowledge hypergraph $G^t = (V^t, \caE^t, \bar{\caE}^t, D)$, where $V^t$ are the vertices, $\caE^t$ are the hyperedges, $\bar{\caE}^t$ are the non-hyperedges, and $D$ are degree informations.
  Namely, $V^t$ consists of all vertices that appeared in queries of the algorithm or in answers to them.
  Similarly $\caE^t$ consists of all hyperedges that were returned for completion queries, and $\bar{\caE}^t$ consists of all sets that are found to be non-existent, that is, completions of completion queries whose answers were $\bot$ or completions of size queries whose answers were $0$.
  Finally, $D$ contains the information on the number of completions of sets of size at most $k$ that were returned by size queries.
  For every hyperedge $S = \set{v_1,\ldots,v_k} \in \caE^t$, the knowledge hypergraph will include the indices of the entries in the tables by which $v_i$'s are connected (that is, $(i_1,\ldots,i_k)$ such that $T_{b_k,B}[v_i, i_1], \ldots,T_{b_k,B}[v_k, i_k]$ are matched, where $B = \set{b_1,\ldots,b_k}$ is a set with $v_i \in V_{b_i}$ for each $i \in [k]$).

  We note that we cannot detect a simplex using size queries (as long as $q(n) = o(n)$).
  Hence, we upper bound the probability that we close a simplex using completion queries.

  \paragraph*{Completion queries of size $k$:}
 Let $B= \set{b_1,\ldots,b_k}$; suppose a completion query $Q = \set{v_1,\ldots,v_k}$ with $v_i  \in V_{b_1}$ for each $i \in [k]$ is given.
  Then, the process computes the probability, conditioned on the current knowledge hypergraph, that $Q$ is a hyperedge.
  Namely, it considers all hypergraphs in the support of $\caD_{n,k}$ that are consistent with $G^t$ and answers positively, that is, $Q$, with probability that is proportional to the number of these hypergraphs that have the  hyperedge $Q$.
  Let $U_{t,b_i,B}$ denote the number of unmatched entries in $T_{b_i,B}$ at time $t$ (before the $(t+1)$-th query).
  Note that this number equals the number of unmatched entries in $T_{b_j,B}$ for every $j\in [k] \setminus \set{i}$ at the same time.

  The probability that there is a hyperedge $Q$ is bounded above by:
  \begin{align}
    \frac{\bar{d} \cdot \bar{d}^{k-1}}{\prod_{i=2}^k\bigl(U_{t,b_i,B }-|\bar{\caE}^{t}|\bar{d}\bigr) }
    \label{eq:prob-size-k}
  \end{align}
  To verify this consider an iterative process in which
  the (at most $\bar{d}$) unmatched entries in $v_1$'s row in $T_{b_1,B}$ are matched one by one to unmatched entries in $T_{b_i,B}$'s for $i \in \set{2,\ldots,k}$.
  After $\tau$ steps, the probability of selecting any of the (at most $\bar{d}$) unmatched entries in $v_i$'s row in $T_{b_i,B}$ for every $i \in \set{2,\ldots,k}$ is the number of combinations of unmatched entries in $v_i$'s row in $T_{b_i,B}$ for $i \in \set{2,\ldots,k}$, divided by $\prod_{i=2}^k(U_{t,b_i,B}- |\bar{\caE}^t| \bar{d})$.
  Here, the term $|\bar{\caE}^t|  \bar{d}$ is the maximum number of unmatched entries that cannot be matched to entries in $v_1$'s row because they correspond to a hyperedge in $\bar{\caE}^t$ containing $v_1$.

  Note that $U_{t,b_i,B} \geq \bar{d} \bar{n} -t$ for every $i \in \set{2,\ldots,k}$.
  We also have $|\bar{\caE}^t|\leq t$ with probability at least $1-o(1)$ (over the choice of the hypergraph), as we get non-hyperedge only when the query is a set of size $k$ from Lemma~\ref{lem:hypergraph-degree}.
  Recalling that $t=o(n)$, the probability~\eqref{eq:prob-size-k} is $O(\bar{d}/\bar{n}^{k-1}) = O(n^{-1/k})$.
  It follows that the probability that the algorithm gets a positive answer, that is, $Q$, for any of the at most $o(n^{1/k})$ queries of size $k$ is $o(1)$.
  But if the algorithm always gets $\bot$ or $0$ for its completion queries of size $k$, it clearly cannot close a simplex with such a query.

  \paragraph*{Completion queries of size at most $k-1$:}

  Suppose that, for $B = \set{b_1,\ldots,b_\ell}$ with $\ell \leq k-1$, a completion query $Q = \set{v_1,\ldots,v_\ell}$ with $v_i \in V_{b_i}$ for each $i \in [\ell]$ is given.
  By Lemma~\ref{lem:hypergraph-degree}, with probability at least $1-o(1)$ (over the choice of the hypergraph), the process answers a completion $S = (v_1,\ldots,v_k)$ of $Q$ according to the conditional probability that the entries $T_{b_1,B}[v_1, i_1],\ldots,T_{b_k,B}[v_k, i_k]$ are matched (where again, the conditioning is on the current knowledge hypergraph $G^t$).
  Here, we would like to upper bound the probability that any of the completed vertices, that is, vertices in $S \setminus Q$, already belongs to the knowledge hypergraph.
  Observe that the knowledge hypergraph $G^t$ contains at most $kt$ vertices and at most $t$ hyperedges and non-hyperedges.

  Suppose $Q = \emptyset$.
  Then, the probability that we hit a vertex in the knowledge hypergraph is upper bounded by
  \begin{align*}
    & \prod_{i=1}^{k}\frac{kt}{\bar{n}-(kt+i-1)-|\bar{\caE}^t|}
    \leq
    {\left(\frac{kt}{\bar{n}-k(t+1)-|\bar{\caE}^t|}\right)}^k=o(1)
  \end{align*}
  as $t = o(n)$.

  Suppose $Q \neq \emptyset$.
  Without loss of generality, we assume $q_1 \in V_{b_1}$ is a vertex and $q_2 = *$.
  Then, the conditional probability that $T_{b_1,B}[v_1, i_1]$ is matched to $T_{b_2,B}[v_2, i_2]$ for some $v_2 \in V^{t-1} \cap V_{b_2}$ and $i_2 \in [\bar{d}]$ at time $t$ is upper bounded by
  \begin{align*}
    \frac{\bar{d}t}{\bar{d}{\bar{n}}-t-|\bar{\caE}^t|\bar{d}}
    \leq \frac{\bar{d}t}{\bar{d}{\bar{n}}-t-t\bar{d}} = o\left(\frac{dn^{1/k}}{dn}\right) = o(n^{-1+1/k}).
  \end{align*}
  as $t=o(n^{1/k})$.

  To summarize, the probability that we hit a vertex in the knowledge hypergraph for any of the at most $o(n^{1/k})$ completion queries of size at msot $k-1$ is $o(1)$.
  But if the algorithm always gets a new vertex for its completion queries of size at most $k-1$, it clearly cannot close a simplex with such a query.
\end{proof}

\end{document}